\documentclass[preprint,12pt]{aastex}

\newcommand{\degree}{\mbox{$^{\circ}$}}
\newcommand{\am}{\mbox{\arcmin}}
\newcommand{\as}{\mbox{\arcsec}}

\def\lsim {$\rlap{\raise.4ex\hbox{$<$}}\lower.55ex\hbox{$\sim$}\,$}

\newcommand{\lsun}{\mbox{L$_\odot$}}
\newcommand{\msun}{\mbox{M$_\odot$}}

\newcommand{\lbol}{\mbox{$L_{bol}$}} 

\input{epsf}

\begin{document}

               
\title {Testing the global star formation relation: An HCO$^+$ (3-2) mapping study of Red MSX sources in the Bolocam Galactic Plane Survey}
\author {David E. Schenck\altaffilmark{1,2} ,
         Yancy L. Shirley\altaffilmark{1,3},
	 Megan Reiter\altaffilmark{1},
         \and St\'ephanie Juneau\altaffilmark{1}
} 

\altaffiltext{1}{Steward Observatory, University of Arizona, 933 Cherry Ave., Tucson, AZ 85721}
\altaffiltext{2}{Senior Honors Thesis}
\altaffiltext{3}{Adjunct Astronomer at the National Radio Astronomy Observatory}

 
\begin{abstract}
We present an analysis of the relation between the star formation rate (SFR) and mass of dense gas in Galactic clumps and nearby galaxies. Using the bolometric luminosity as a measure of SFR and the molecular line luminosity of HCO$^+$ (3-2) as a measure of dense gas mass, we find that the relation between SFR and M$_{dense}$ is approximately linear. This is similar to published results derived using HCN (1-0) as a dense gas tracer. HCO$^+$ (3-2) and HCN (1-0) have similar conditions for excitation. Our work includes 16 Galactic clumps that are in both the Bolocam Galactic Plane Survey and the Red MSX Survey, 27 water maser sources from the literature, and the aforementioned HCN (1-0) data. Our results agree qualitatively with predictions of recent theoretical models which state that the nature of the relation should depend on how the critical density of the tracer compares with the mean density of the gas.

\end{abstract}

\keywords{Galaxies: star formation-Radio lines: general-Stars: formation}


\section{Introduction}

All stars originate in clouds of cold, dense molecular gas (Evans 1999). The formation of low-mass stars has been studied in depth and the process by which it occurs is believed to be fairly well understood (Shu, Adams, \& Lizano 1987, Andr\'e et al. 2000). However, the formation of stars with masses greater than $\sim$8 $\msun$ is more difficult to examine because the regions in which they form are considerably farther away (Wu et al. 2010). High-mass stars may not simply be a scaled-up version of low-mass stars, but instead may be formed in a fundamentally different way (e.g., competitive accretion, core coalescence vs. scaled-up accretion; see McKee \& Ostriker 2007). A complete theory of star formation requires an understanding of the efficiency and rate of forming stars across the entire stellar mass range.

In the last few years, a new nomenclature has been adopted to describe the structure of Galactic star-forming regions within molecular clouds.  Protostars form from individual dense cores of gas.  Most protostars form in clustered environments and collections of cores which may be physically connected or gravitationally bound within a larger clump.  Clumps located in the Milky Way are near enough to be observed in detail using submillimeter single-dish telescopes with resolutions of tens of arcseconds (e.g., Reiter et al. 2011). The information gained from studies of these local star-forming clumps can be used to understand star formation in galaxies too far away to resolve individual star-forming regions. Ultimately, these studies may be used to constrain a universal star formation law.

Determining the rate and efficiency of star formation has been a long standing problem. Schmidt (1959) first proposed that the rate of star formation would depend on the volume density of dense gas according to a power law. Building upon this assumption, it was shown globally for galaxies that the power law phrased in terms of surface density gave $\Sigma_{SFR} \propto \Sigma_{gas}^{1.4 \pm 0.15}$ (Kennicutt 1998). This relation, which was derived using CO to trace the molecular gas and HI to trace the atomic gas is called the Kennicutt-Schmidt relation (K-S relation).  Even with the resolution of state-of-the-art interferometers, the K-S relation has only been probed down to scales of $\approx 1$ kpc, much larger than the sizes of giant molecular clouds (tens of pc) in nearby galaxies (e.g., Bigiel et al. 2011).  The K-S relation has been questioned because CO observations probe low density molecular emission and not the dense molecular gas more directly associated with the 
clumps and cores from which protostars form (e.g. Gao \& Solomon 2004a).
Therefore, subsequent studies have tested the power law relation by measuring the molecular line luminosity of tracers that specifically trace the dense gas mass. Infrared luminosity is used as a measure of the star formation rate; collapsing protostars emit strongly in the IR. Using the line luminosity of the transition HCN (1-0) and infrared luminosity of a set of 65 galaxies, including luminous and ultraluminous infrared galaxies (LIRGs and ULIRGs), Gao \& Solomon (2004a,b) found a power law index of $1.00 \pm 0.05$.  They interpreted this result as the SFR being proportional to dense gas mass (Gao \& Solomon 2004a,b).  This HCN (1-0) survey was extended to the scale of Galactic dense clumps with similar results; the power law index was very near one over a range of 7 orders of magnitude in $L_{IR}$ (Wu et al. 2005, Wu et al. 2010). This was interpreted as evidence that there is a fundamental unit of star formation associated with massive clumps with luminosities L$_{bol} > 10^{4.5}$\lsun\ and the star formation efficiency within the dense clumps is constant. 


Since the initial work of Gao \& Solomon and Wu et al., a wider range of critical densities has been probed by different molecular tracers (e.g., Bussmann et al. 2008, Juneau et al. 2009). Theoretical radiative transfer models predict the behavior of the SFR and line luminosity as a function of gas density on galactic scales (Krumholz and Thompson 2007, Narayanan et al. 2008). The behavior of the line luminosity is dependent on the critical density of the tracer in these models; if the critical density exceeds the mean density, the line luminosity rises superlinearly with density. However, for critical densities below the mean density, it rises linearly. The SFR is assumed to rise superlinearly with density in these models \footnote{$SFR \propto n^{1.5}$, where $n$ is the density of molecular hydrogen in cm$^{-3}$}, similar to the original Schmidt relation (1959) but different from the conclusion of Gao \& Solomon. Combining the two conditions gives linear correlations for high critical density tracers and superlinear correlations for low critical density tracers. This general trend is observed in galaxies using CO (1-0), HCN (1-0) and (3-2), and HCO$^+$ (1-0) and (3-2) (Juneau et al. 2009, also see Baan et al. 2008, Iono et al. 2009).

Gao and Solomon (2004) and Wu et al. (2010) interpret their results as evidence that the SFR to dense gas mass ratio is constant. However, the extragalactic community and many theoretical models continue to assume a superlinear Schmidt relation. In this paper, we test the observed Gao \& Solomon and Wu et al. correlation by mapping a sample of 16 massive clumps in HCO$^+$ (3-2) selected from the Bolocam Galactic Plane Survey (Aguirre et al. 2011). This sample is combined with 27 massive clumps mapped in HCO$^+$ (3-2) that are associated with water masers from Reiter et al. (2011) and 14 (U)LIRGs, from Juneau et al. (2009) that were adapted from Graci\'a-Carpio et al. (2008) using updated distances. We also re-analyze the 50 dense clumps from Wu et al. (2010) and 42 galaxies from Gao and Solomon (2004) to verify the published HCN correlation.

We choose HCO$^+$ (3-2) because of its ability to trace nearly identical excitation conditions to HCN (1-0). The frequency and dipole moment of HCO$^+$ (3-2) are a factor of $\sim$3 and $\sim$1.5 times higher, respectively, than the corresponding values of HCN (1-0). The Einstein A coefficient is proportional to $\nu^3$$\mu^2$, making the Einstein A coefficient of HCO$^+$ (3-2) $\sim$  60 times higher. Most of this disparity is compensated by HCO$^+$ inducing dipoles in the molecules with which it interacts, increasing the effective cross section for collision and the corresponding collision rate $\gamma_{jk}$. HCN is not an ion, so it does not induce a dipole in nearby molecules and it has a comparatively lower rate of de-excitation. At 20 K, HCO$^+$ (3-2) and HCN (1-0) have critical densities of $3.90 \times 10^6$ and $1.25 \times 10^6$ cm$^{-3}$, respectively, a factor of $\sim$3.1 difference. 
Since the excitation of a molecular line varies gradually over a range of densities, claiming the line traces gas at the critical density is an oversimplification (Evans 1999). A more practical quantity is the effective density for excitation, $n_{eff}$, which is defined as the density at which a transition will have a radiation temperature of 1 K assuming log(N/$\Delta$v)=13.5 cm$^{-2}$/km/s and an average kinetic temperature of 20 K (Evans 1999, Reiter et al. 2011).
The two transitions also have even more similar effective densities;  HCO$^+$ (3-2) and HCN (1-0) have effective densities of $2.33 \times 10^4$ and $1.50 \times 10^4$ cm$^{-3}$. These only differ by a factor of $\sim$1.5. Densities can range over several orders of magnitude within molecular clouds, so such a small difference means that the transitions essentially trace the same molecular gas in the absence of strong chemical differentiation.

In \S2, we summarize the observations of our sample of 16 massive cores as well as the full (combined) sample used in this paper. In \S3, we discuss the methods used to derive physical quantities from the maps. We present our results and compare them to both other observations and theoretical predictions in \S4 as well as critically discuss several caveats with the interpretation of the observed correlation. 

\section{Observations}
Sixteen sources were mapped using the 10-meter Heinrich Hertz Telescope (HHT) on Mount Graham, Arizona. Observations were made on April 10 and 11 as well as May 4 through May 9, 2010, excluding May 5. The HHT was equipped with the 1mm ALMA prototype sideband-separating receiver. The observations were made with a 6.0 GHz IF in dual polarization 4 IF mode with the vertical and horizontal polarizations each split into upper and lower sidebands. Two transitions were observed simultaneously; HCO$^+$ J $= 3 \rightarrow 2$ (267.558 GHz) was centered in the lower sideband while N$_2$H$^+$ J $= 3 \rightarrow 2$ (279.512 GHz) was centered in the upper sideband. The N$_2$H$^+$ maps are not included in this analysis because of the weakness of the detections. The observed line properties are listed in Tables 2 and 3. The backend filter-banks were split into 256 channels with 250 kHz resolution. At the frequencies used, the velocity resolution was $\sim$0.28 km s$^{-1}$. The maps have an angular resolution of $27.2^{\prime\prime}$.

Maps were made using the On-The-Fly (OTF) imaging technique (Mangum et al. 2007). Scan rows were separated by about 10{\arcsec}, about one-third of the beam size. Each source was scanned at least once in the RA and DEC directions and many were scanned again in both. Additional pairs of scans were offset by about 5{\arcsec} from the first pair to better sample the source. The pointing accuracy of the HHT is typically 5{\arcsec} RMS.

The maps were reduced using the GILDAS CLASS software. First, a linear baseline was removed from each map. The vertical and horizontal polarizations were scaled separately using their respective main beam efficiencies and then combined ($\eta_H$=0.70$\pm$0.04, $\eta_V$=0.81$\pm$0.04). The maps were convolved using a Gaussian-tapered Bessel function of the form $\frac{J_1(\frac{r}{a})}{(\frac{r}{a})}exp[-(\frac{r}{b})^2]$ where a = 1.55$(\frac{\theta_{mb}}{3})$ and b = 2.52$(\frac{\theta_{mb}}{3})$ (Mangum et al. 2007, Reiter et al. 2011) to preserve spatial resolution.

\subsection{Source Selection}
The sixteen sources mapped for this project were selected from the Bolocam Galactic Plane Survey (BGPS). A total of 1402 BGPS sources spectroscopically detected by Schlingman et al. (2011) were examined by eye using GLIMPSE (8 $\mu$m) and MIPSGAL (24 $\mu$m) images. The MIPSGAL images were used to determine which sources were associated with a 24 $\mu$m point source. The point sources indicate the presence of either an embedded protostar or an evolved star. Each BGPS source within 15{\arcsec} of the point source was considered to contain a candidate embedded object. The 247 objects associated with 24 $\mu$m point sources were then compared to the Red MSX Source (RMS) survey by calculating the offset between their positions. The RMS survey includes over one thousand massive young stellar objects (MYSOs) that are radio-quiet, mid-infrared point sources (Mottram et al. 2010). Comparing the coordinates of the objects in each catalogue, 79 from MSX were within 15$\arcsec$ 
of a BGPS source observed by Schlingman et al. The Schlingman et al. subsample of BGPS sources contains the highest flux 1.1 mm sources in the BGPS.
Candidate RMS sources in the BGPS were observed generally in order of brightest to dimmest HCO$^+$ (3-2) intensity. Not every object common to the two surveys was observed due to observational constraints, but every joint object with I$_{HCO^+}$(3-2) $>$ 10.5 K km s$^{-1}$ was mapped.

In addition to our observations, we include HCO$^+$ (3-2) maps from the survey of Reiter et al. (2011).  Twenty-seven massive clumps associated with H$_2$O masers were taken from this paper. The luminosity and distance of the BGPS-RMS sources and water maser sources are listed in Table 1. The bolometric luminosity is given for the BGPS-RMS sources while the infrared luminosity is given for the water maser sources. Also, fourteen luminous and ultraluminous infrared galaxies (LIRGs and ULIRGs) originally from Graci{\'a}-Carpio et al. (2008) represent the currently known extragalactic observations of HCO$^+$ (3-2). We analyze the correlation between bolometric luminosity and molecular line luminosity for this combined sample using a Bayesian linear regression technique (Kelly 2007). We also re-analyze the Wu et al. (2010) correlation of 50 massive, dense clumps and 42 galaxies that were mapped in HCN (1-0)(Wu et al. 2010, Gao and Solomon 2004a,b) using the same Bayesian linear regression technique. This was done in order to characterize the uncertainty in the slope of the correlation as well as make the correlations of the two tracers more directly comparable. Galaxies which had upper or lower limits on the molecular line luminosity were not used in our analysis.

\section{Clump Physical properties}

The maps in integrated intensity of HCO$^+$ (3-2) emission are shown in Figure 1.
Integrated intensities are given by,
\begin{equation}
I(T_{mb})=\int T_{mb} dv \pm \sqrt{\delta v_{line} \delta v_{chan}} {\sigma}_{T_{mb}}
\end{equation}
where $\delta v_{line}$ is the velocity extent of the line, $\delta v_{chan}$ is the channel width of the spectrometer (0.28 km s$^{-1}$), and $\sigma_{T_{mb}}$ is the uncertainty in the main-beam temperature. The quantity $\delta v_{line}$ was derived from setting boundaries on each side of the line so that the entire line was enclosed. As a result, the intensity uncertainty is a conservative estimate. The line with the highest integrated intensity in each map was used in calculating other quantities. The peak integrated intensities are usually near the center of the map, but that is not always true, especially when there are two objects in a single map. The total sample of galactic cores mapped in HCO$^+$ (3-2) had a mean intensity of $52.6 \pm 50.2$ K km$^{-1}$ and a median of 36.5. The standard deviation of the sample is relatively large because the sources cover a wide range of intensity from 3.2 to 195.9 K km s$^{-1}$. The mean and median for the subset of BGPS/RMS selected cores are $13.6 \pm 9.5$ K km s$^{-1}$ and 11.7 K km s$^{-1}$. The mean and median for the subset of water maser selected sources are $75.6 \pm 50.3$ K km s$^{-1}$ and 61.5 K km s$^{-1}$ (Reiter et al. 2011). The BGPS selected subset extends the Reiter et al. sample of Galactic sources to lower integrated intensities in HCO$^+$ (3-2).

The angular diameter was chosen to be the area within the the half-peak intensity contour, A ($\theta_{s*b} = \sqrt{4A/\pi}$). The uncertainty in the half-peak diameter was determined by finding the diameters associated with the half-peak contour $\pm$ $\sigma_{I(T_{mb})}$. The beam of the telescope (27.2{\arcsec}) contributed to the width of the maps. This effect was removed by calculating the deconvolved sizes using
\begin{equation}
{\theta}_{dec}=\sqrt{{\theta}_{s*b}^{2}-{\theta}_{beam}^{2}}
\end{equation}
The distance and angular diameter were used to calculate a physical radius from
\begin{equation}
R=\frac{\theta_{dec} D}{2}
\end{equation}
where R is the radius in the units of the distance, D, and $\theta_{dec}$ is the deconvolved angular diameter in radians. The total sample has a mean size 0.31 $\pm$ 0.18 pc and a median of 0.28 pc. The means of the BGPS/RMS and water maser samples are 0.35 $\pm$ 0.19 and 0.29 $\pm$ 0.16 pc and medians 0.34 and 0.24 pc, respectively. The BGPS/RMS mean is skewed upward by a large source (019.884-053) which is also the brightest in that sample. A histogram of the sizes is presented in Figure 2 along with histograms of other physical properties.


Distances were obtained by using radial velocity measurements to get kinematic distances (Schlingman et al. 2011). Kinematic distances are ambiguous in that two different distances correspond to the same radial velocity in the first quadrant of the Galaxy. Four sources were assigned to the near distance because they appear to be associated with an infrared dark cloud (IRDC) as determined by visual inspection of MIPSGAL images.  The remaining distances were determined using information supplied by the RMS online catalog (Urquhart et al. 2008) which not only gives kinematic distances, but also a "complex distance." The complex distance was for any MYSOs belonging to a complex with a previously determined distance. If one of the two kinematic distances was within about 1 kpc of the quoted complex distance, that kinematic distance was assigned. The water maser clumps from Reiter et al. (2011) used distances that were previously determined (see Shirley et al. 2003).

The uncertainty in the distances is needed to propagate the uncertainty in the molecular line luminosity. The uncertainty was determined using distance probability curves produced by Ellsworth-Bowers et al. (in prep.). This method employs radial velocity measurements, proximity to IRDCs, H I self-absorption, and other factors to calculate a joint probability distribution (P(D)). The distance probability curves are typically bi-modal (centered on the near and far kinematic distances). Each mode of the probability curves were fit by Gaussians, one for the near distance and one for the far distance, to determine the spread of the curves.  Probability curves were available for all of the BGPS/RMS crossover sources, but only 8 of the 27 water maser sources (Ellsworth-Bowers, private communication 2010). The remaining 19 sources without distance probability curves were assigned the average uncertainty of those 8. There was no trend between distance and distance uncertainty.

The bolometric luminosities were calculated using the observed spectral energy distributions (SED) of each source. SEDs were constructed uniformly from fluxes taken from the BGPS (1.1 mm), MSX (8, 12, 14, and 21 $\mu$m), SCUBA (850 $\mu$m), MIPSGAL (70 $\mu$m), IRAS (12, 25, 60, and 100 $\mu$m), and AKARI (65, 90, 140, and 160 $\mu$m). Bolometric luminosities were calculated by integrating over the observed fluxes. Luminosities were already available for the water maser sources, but they were re-calculated because new fluxes had become available since the published SEDs in Mueller et al. (2002).  See Table 4 for the complete SEDs of the sources. The logarithm of the mean luminosity (in $\lsun$) of the sample is $5.09 \pm 0.87$ with a median of 4.52. The mean bolometric luminosity of the water-maser sources exceeds that of the BGPS/RMS sources; the former is $5.23 \pm 0.76$ and the latter $4.64 \pm 0.66$. 

$L^{\prime}$, or the molecular line luminosity, is a commonly used quantity in extragalactic studies as an analogue to the mass of molecular gas. Formally, molecular line luminosity, or source-integrated surface brightness as it is called in Mangum et al. 2008, is defined as
\begin{equation}
L'=23.504\pi \theta_{s*b}^2 (1+z)^{-3} D_L^2 I(T_{mb}) 
\end{equation}
where $\theta_{s*b}$ is the convolved angular diameter in arcseconds, $z$ is the redshift, $D_L$ is the luminosity distance in Mpc, and $I(T_{mb})$ is the integrated intensity in K km s$^{-1}$. Two sources with the same main beam brightness temperature and spatial extent will have the same L$'$. The mean of the line luminosity for the BGPS/RMS and water-maser samples were each 33.5 K km s$^{-1}$. Histograms of the bolometric and line luminosities are shown in Figure 3. In \S4.1 we construct the correlation between bolometric luminosity and molecular line luminosity with distance uncertainties propagated for the Galactic sources.

Finally, we calculated the mass in the Galactic clumps using two different methods.
Virial masses were calculated using
\begin{equation}
M_{vir}=\frac{5 R (\Delta v)^2}{8 a_1 a_2 G ln(2)} \approx \frac{209 R (\Delta v)^2}{a_1 a_2}
\end{equation}
\begin{equation}
a_1=\frac{1-p/3}{1-2p/5}
\end{equation}
where R is the physical radius in pc, $\Delta v$ is the FWHM linewidth in km s$^{-1}$, and $a_1$ and $a_2$ are factors that account for a power law density distribution (n $\propto$ $r^{-p}$) and non-spherical shape, respectively (Bertoldi and McKee 1992). The median value of $p$ (1.75), the exponent of the density power law, was adopted from Mueller et al. (2002), making $a_1$=1.39 for the BGPS/RMS crossover sources. The values of $a_1$ were determined in Mueller et al. (2002) for the water-maser sources. The correction factor $a_2$ is negligible for aspect ratios below 2, so it was set to 1 for all calculations. The linewidth is a measure of the thermal and turbulent support against gravitational collapse, but can be broadened by optically thick emission, leading to overestimates of the virial mass (Shirley et al. 2008, Reiter et al. 2011). The logarithm of the mean virial mass (in $\msun$) is $3.30 \pm 0.41$ and a median of $3.13$. For the BGPS/RMS and water-maser samples,
 the mean is $3.24 \pm 0.41$ and $3.33 \pm 0.41$, respectively.

Masses were also estimated from the 1.1 mm continuum emission by assuming a single dust temperature and common dust opacity for all sources. The isothermal mass is
\begin{equation}
M_{iso}=\frac{S_{\nu} D^2}{B_{\nu}(T)\kappa_{\nu}}
\end{equation}
where S$_{\nu}$ is the total integrated flux at 1.1 mm and $\kappa_{\nu}$ is the opacity of that wavelength. The value $\kappa_{1.1mm}$=0.0114 cm$^2$ g$^{-1}$ was adopted (Ossenkopf \& Henning 1994, Enoch et al. 2006, Battersby et al. 2010), consistent with a dust-to-mass ratio of 100. Assuming a temperature of 15 K, the temperature used in Battersby et al. (2010), the equation for M$_{iso}$ becomes
\begin{equation}
M_{iso}=14.32(e^{13/T}-1)\left(\frac{S_{\nu}}{1 Jy}\right)\left(\frac{D}{1 kpc}\right)^2 \msun.
\end{equation}
The logarithm of the mean isothermal mass (in $\msun$) is $3.37 \pm 0.64$. This is higher than the mean for the virial mass. The water-maser sample also has a higher mean isothermal mass ($3.54 \pm 0.51$), but the mean isothermal mass is lower than the mean virial mass for the BGPS/RMS sources ($3.09 \pm 0.87$).

\section{Results and discussion}
A Bayesian linear regression routine from Kelly (2007) was used to test the correlation between L$'$ and $L_{bol}$ as well as L$'$ and mass. The program uses a Markov chain Monte Carlo (MCMC) method to take random draws of the slope, intercept, and intrinsic scatter by perturbing the previous set of line parameters. These new sets of parameters are either accepted or rejected according to the Metropolis-Hastings algorithm (Metropolis et al. 1953, Hastings 1970). The program tests 10,000 sets of parameters and saves the distribution of accepted values. The spread in the distribution of the slopes is entirely attributable to the uncertainties in the data; the uncertainty from the MCMC sampler is negligible (Brandon Kelly, private communication 2011). For a more detailed description of the program, see Kelly (2007). 

The results of the Bayesian linear regression fits are presented in Table 5. The lines take the form
\begin{equation}
\log(y) = b \times \log(x) + a.
\end{equation}
The ordinate and abscissa variables are listed in Table 5 in the order y-x. 
The quantity $\sigma_{int}$ is the intrinsic scatter of the fit.

\subsection{M-L$'$ correlation}
First we test the correlation between mass and line luminosity.
The masses of Galactic clumps were found using both isothermal mass (determined from dust continuum emission) and virial mass (determined from HCO$^+$ emission). The M$_{iso}$-L$'$ relation (Figure 3(a)) had a sublinear slope ($0.78 \pm 0.10$) and the M$_{vir}$-L$'$ relation (Figure 3(b)) had a linear slope ($1.06 \pm 0.18$). 
Using M$_{iso}$ produced a slightly tighter correlation; the correlation coefficient for M$_{iso}$ was 0.85 compared to only 0.70 for M$_{vir}$. 
The same trends were observed in multiple molecular species in Reiter et al. (2011) and Wu et al. (2010) using the water maser sources.

The virial mass depends on the linewidth, $\Delta v$. The line can be artificially broadened if the emission is optically thick which is true for the transitions considered (see Phillips et al. 1979). This effect will cause the virial mass to be an overestimate. M$_{iso}$, on the other hand, does not depend on the line parameters, unlike M$_{vir}$ and $L^{\prime}$. Thus, we prefer M$_{iso}$ as it is an independent estimate of the clump mass. Nevertheless, our tight linear correlation with both mass estimates indicates that $L^{\prime}$ is indeed a good tracer of the dense gas in clumps.

\subsection{L$'$-L$_{bol}$ correlation}

Correlations were found for several subsets of the data by using L$'$ as the independent variable and L$_{bol}$ as the dependent variable. These are presented in Figure 3(c) (HCO$^+$ (3-2)) and Figure 3(d) (HCN (1-0)).
The slope of the galactic clumps and combined data sets observed in HCO$^+$ (3-2) both came out to nearly one ($1.10 \pm 0.12$ and $1.04 \pm 0.02$ respectively). The histograms in the bottom right corner of each panel in Figure 3 show the distribution of slopes for the galactic and extragalactic data.
The slope for all clumps and galaxies together has an uncertainty nearly a factor of 10 lower than the clumps alone. This is the result of the data appearing in two widely separated groups in L$_{bol}$-L$'$ space so that the line was basically fit to two clumps of points. 
The slope of the correlation for Galactic clumps agrees very well with the Wu et al. (2010) correlation for HCN (1-0) (also see below). While the best-fit line for the Galactic clumps extends to be slightly higher than the extragalactic objects, the uncertainty on the slope makes it such that the extragalactic points are within 1$\sigma$ of the fit.  The extrapolation of the best-fit of the extragalactic points misses the Galactic points more dramatically, although the large uncertainty on the slope still cannot rule out a linear slope at the more than $1\sigma$ level.  This result is different than was found for HCN (1-0) by Wu et al. (2005),  however, we must caution that the published extragalactic HCO$^+$ (3-2) detections are small in number and limited to $L_{bol} > 10^{11}$ $\lsun$. Global extragalactic observations of L$^{\prime}$ in HCO$^+$ (3-2) are needed, especially for galaxies with L$_{bol}$ $<$ 10$^{11}$ $\lsun$ to better compare with the HCN (1-0) results.  

Linear fits for the HCN (1-0) data set were presented in Gao and Solomon (2004a,b) and Wu et al. (2010), both of which found a linear relation between L$_{IR}$ and L$'$. We test their results by re-analyzing their data with the Bayesian routine. The line luminosity of one source, DR21S, was misprinted in Wu et al. (2010) (it is 2.29 rather than 0.29; Wu, private communication 2011). This did not affect their results as their correlations were found using the correct value. The fits displayed in the top right plot of Figure 2 do not include any points below L$_{IR}$=10$^3$ $\lsun$. This cutoff was tested because that is roughly the lowest L$_{bol}$ in the HCO$^+$ data set. The limit makes the HCN (1-0) and HCO$^+$ (3-2) correlations span the same range in luminosity. The slopes for the galactic, extragalactic, and combined sets are $1.17 \pm 0.11$, $1.07 \pm 0.06$, and $0.99 \pm 0.01$ respectively. These slopes are all within 1.5 $\sigma$ of unity. The agreement between the extrapolated 
galactic and extragalactic best-fit lines is better than for HCO$^+$ (3-2). The extragalactic correlation from Gao \& Solomon (2004a,b) spans $1.5$ more orders of magnitude in $L_{bol}$ than the observed HCO$^+$ (3-2) in galaxies and likely provides a more robust sample from which to determine the fit. Our re-analysis of the Gao \& Solomon and Wu et al. data confirms their results.

The amount of dense gas in the galactic clumps needs to be high enough to fully sample the IMF. The IMF describes the number of stars per mass bin and, generally, lower mass stars are more common. A cloud with a relatively small mass will not form as many stars which makes the probability of seeing a massive star lower. When we say "sample the IMF", we mean the probability of seeing stars across the full range of the IMF is high. Based on the assumption that larger clouds with more stars will be more luminous, Wu et al. (2010) instituted a minimum L$_{IR}$ of 10$^{4.5}$ $\lsun$. The linear regression was performed with this condition imposed in order to directly compare to their work. The slopes were consistently lower; the galactic and combined samples had slopes 0.88 $\pm$ 0.15 and 0.94 $\pm$ 0.02, respectively. These slopes may be smaller because of the way the cutoff was imposed. Simply 
removing all points below a certain luminosity will allow more scatter in the horizontal direction compared to the vertical which will flatten the line.

In this work, we choose to define the cutoff as a minimum line luminosity rather than infrared or bolometric luminosity. The motivation behind the cutoff is to define a minimum mass which we have constrained from the observed correlations between line luminosity and mass. Krumholz and Thompson (2007) claim that the cloud mass should exceed $\sim$1000 $\msun$ to accomplish a decent sampling. Using the correlation between L$'$ and M$_{iso}$, setting the mass to 1000 $\msun$ gives a line luminosity of 0.85 K km s$^{-1}$. Based on the Galactic L$_{bol}$-L$'$ correlations with no cutoff, this corresponds to bolometric (or infrared) luminosities of 10$^{4.2}$ and 10$^{3.4}$ for HCO$^+$ and HCN, respectively. The region containing points that are below both cutoffs are indicated by dotted lines in the bottom left corner of Figure 3(c) and 3(d). In both cases, most of the points in these regions lie below the correlation, indicating they may not be sampling the IMF well enough. There
 are not many points in these regions and the downturn observed in HCN (1-0) is not apparent in HCO$^+$ (3-2), so it is not possible to definitively conclude where the cutoff should be, but there is no evidence against those derived assuming 1000 $\msun$ as the minimum mass necessary.

\subsection{Comparison to theory}

The physical basis for the observed L$'$-L$_{bol}$ correlations have been studied using numerical simulations. Krumholz and Thompson (2007) modeled isothermal and homogeneous giant molecular clouds (GMCs) using a radiative transfer code and considering star formation regulated by turbulence. Narayanan et al. (2008) used 3D non-LTE radiative transfer with hydrodynamic simulations of isolated galaxies and galaxy mergers. Despite considering two different classes of objects, the same explanation was reached by both. The origin of the SFR-L$'$ relation comes from the Schmidt relationship and the density dependence of L$'$. The Schmidt relation says that the star formation rate is proportional to density to some power, or
\begin{equation}
SFR \propto \rho^N
\end{equation}
where $N$ is the Schmidt index. The value of $N$ is 1.5 with one factor of $\rho$ coming from the mass available for stars and a factor of $\rho^{0.5}$ coming from the density dependence of the dynamical time for collapse (Madore 1977, Elmegreen 1994). This assumes that star formation occurs at a constant rate per dynamical time. To predict the relation between SFR and line luminosity, 
\begin{equation}
SFR \propto \L'^{\alpha} ,
\end{equation}
the dependence of the line luminosity on density must be known. This is expressed as
\begin{equation}
L' \propto \rho^{\beta}
\end{equation}
so that the index of the SFR-L$'$ correlation is $\alpha$=$\frac{N}{\beta}$. 

Both simulations determined that density dependence of L$'$ fell into two regimes, one in which the critical density is much larger than the mean density of the gas and one in which it is much smaller. When n$_{crit}$ $>>$ $\bar{n}$, only the high density tail of the density distribution is above the critical density so emission predominantly arises from the densest regions. The emission does not trace the density of the gas, but instead will trace the same density in every subcritical object. As density increases (with n$_{crit}$ remaining much larger than $\bar{n}$), the line emission rises superlinearly so that $\beta$ $>$ 1 and $\alpha$ $<$ N. The SFR-L$'$ relation will have an index below the Schmidt index. Both HCO$^+$ (3-2) and HCN (1-0) fall in this regime. In the other case where n$_{crit}$ $<<$ $\bar{n}$, nearly all of the gas is thermalized and emitting. The line luminosity faithfully traces the mass and rises linearly with density. The index of the SFR-L$'$ relation will approximately equal the Schmidt index. CO (1-0) produces an index near the Schmidt index, which is consistent with this formulation as CO traces lower densities than HCO$^+$ and HCN.

Some combination of gas conditions and molecular transitions will be intermediate between the two cases. As density increases, the relation passes from the subcritical to the supercritical case, meaning the SFR-L$'$ relation should transition from linear to superlinear. Evidence of this upturn have been found in observations, though it is only tentative due to a caveat to be discussed in the next subsection (Gao et al. 2007).

A power law index of 1.0 was predicted for HCN (1-0), in agreement with observations. The critical density of HCO$^+$ (3-2) is about a factor of 3 higher than that of HCN (1-0). Gas tracers have critical densities that span several orders of magnitude, so the difference in n$_{crit}$ for the two tracers considered here is relatively small and they can be considered to trace the same conditions. Based on the theoretical predictions, the index for HCO$^+$ (3-2) should also be near 1.0. 
The results presented in this paper are consistent with the theoretical predictions of Krumholz \& Thompson and Narayanan et al. The interpretation that a linear relation means a constant ratio between the SFR and dense gas mass cannot be distinguished from the theoretical predictions based on our work.

\subsection{Caveats}

The L$_{bol}$-L$'$ relation is intended to represent a relation between the SFR and gas mass. However, the conversions are complicated by a wide range of assumptions which must be explored. 

Bolometric (or infrared) luminosity is the most popular measure of the star formation rate (SFR) as it is easy to determine. On scales from 1 kpc to galactic scales, the IMF of stars is well sampled and the conversion from $L_{bol}$ to SFR is linear (see Kennicutt 1998).  However, this conversion for galaxies is potentially confounded by the presence of active galactic nuclei (AGN) which contribute their own IR emission by heating dust grains (Andreani et al. 2003, Carilli et al. 2005, Li et al. 2008). The extra observed IR flux can lead to an overestimate of the SFR.

The effect of AGN on the L$_{bol}$-L$'$ correlation is expected to be negligible. For a sample of 68 ULIRGs, about 85 $\%$ of the infrared luminosity was found to be powered by star formation with only 15 $\%$ attributable to AGN (Nardini et al. 2008). While AGN occur more frequently in more IR-luminous galaxies (e.g., Veilleux et al. 1995), their fractional contribution to the total bolometric luminosity mostly remain small. The net effect on the L$_{bol}$-L$'$ relation was tested in Juneau et al. (2009) by comparing the slopes obtained when using either the total infrared luminosity (L$_{IR}$) or the far-infrared luminosity (L$_{FIR}$) as a proxy for L$_{bol}$. The far-infrared luminosity is expected to be dominated by star formation even in the presence of AGN-heated dust emission, which peak in the mid-infrared. In contrast, the total infrared emission is more subject to AGN contamination. However, the L$_{bol}$-L$'$ relation is unchanged when substituting L$_{IR}$ for L$_{FIR}$, suggesting that possible AGN contributions do not affect the relation (at least within the uncertainties).

While Galactic clumps will not be affected by AGN, their $L_{bol}$ also is not a perfect measure of the SFR. The most massive stars in a system younger than $\sim$3 Myr will not have had time to evolve and die in a supernova; therefore, not enough time has elapsed to achieve equilibrium between stellar birth and death. In this situation, the light emitted will not trace the star formation rate. A simulation run by Krumholz and Tan (2007) found that the luminosity per unit star formation is insensitive to the age for populations older than a few Myr (which is a timescale that is easily averaged over 1 kpc within a galaxy but would not necessarily apply on the scale of an individual Galactic clump). The luminosity of younger populations traced mass, but not the star formation rate. Dividing the mass by an independent measure of the age to estimate the star formation rate does not solve the problem for systems younger than $\sim$1 Myr because not all of the stars in such a system have finished collapsing. These pre-main sequence stars are powered by accretion and will release more energy than a main sequence star of the same mass. Since the contribution of these collapsing objects is unknown, it is impossible to determine how much mass is in the system (Krumholz and Tan 2007).  By resolving the infrared population in Galactic clumps (e.g., with James Webb Space Telescope or Large Binocular Telescope Interferometer) and directly comparing the mass distribution within the clumps (e.g. with ALMA), we may be able to better constrain the actual SFR in Galactic clumps.

Chemical variations potentially make the utility of HCN and HCO$^+$ questionable as uniform dense gas tracers in different environments. The strong X-ray emission from an AGN can increase ionization and therefore the abundance of HCN in relation to other molecules (Lintott \& Viti 2006). This is the explanation given for the trend observed in the HCN(1-0)/HCO$^+$(1-0) luminosity ratio which increases with increasing far-infrared luminosity for L$_{FIR}$ $>$ 10$^{11}$ $\lsun$ by Graci{\'a}-Carpio et al. (2008). However, another interpretation is that more luminous galaxies have a higher dense gas fraction and the effects on HCN chemistry are negligible. Using 5 transitions from 3 molecules (HCN, HCO$^+$, CO) to form 10 line luminosity ratios (with the higher critical density tracer in the numerator), Juneau et al. (2009) showed that the ratios consistently increased with L$_{IR}$ except for the HCO$^+$(3-2)/HCN(1-0) ratio which combines the two tracers considered here. This ratio, which has tracers of nearly the same critical density, was approximately constant. Using the same simulation as Narayanan et al. (2008), the line ratio behavior was replicated without including chemical abundance changes. They concluded the dense gas fraction increased with luminosity and that inclusion of chemical effects in the dense gas were not necessary.


Molecular gas tracers are supposed to indicate the presence of molecular gas through collisions, emitting more energy when they undergo more collisions. Ideally, all emission would be a result of interactions with the gas, but the emission can be affected by the background radiation. The excitation of HCN lines may be enhanced by IR pumping of a 14 $\mu$m vibrational transition (Graci{\'a}-Carpio et al. 2006). Similarly, HCO$^+$ excitation can be affected by a 12 $\mu$m vibrational transition. Any emission associated with radiative pumping will lead overestimates of the gas mass inferred from the line luminosity.

The complications discussed above present real obstacles to interpreting the L$_{bol}$-L$'$ relation. None of them were included in the GMC or galaxy radiative transfer modeling because of the high degree of complexity that each would require. It is argued in both Krumholz and Thompson (2007) and Narayanan et al. (2008) that their ability to replicate the relations for different tracers without accounting for these details shows that their effects may be negligible.
Despite these caveats, the results from this study show similar trends for the L$_{bol}$-L$'$ relation of HCO$^+$ (3-2) as was observed by Wu et al. (2005, 2010) for HCN (1-0).  

\section{Summary}

We conclude that line luminosity in HCO$^+$ (3-2) is tracing mass as evidenced by the fits of L$'$ with isothermal mass and virial mass. M$_{vir}$ rises roughly linearly with L$'$ while M$_{iso}$ also increases with L$'$, but with a sub-linear slope.

The slope of the log(L$_{bol}$)-log(L$'$) relation for HCO$^+$ (3-2) is 1.04 $\pm$ 0.02 (1.09 $\pm$ 0.12 for Galactic clumps and 0.81 $\pm$ 0.21 for galaxies). This would imply a linear relation between the star formation rate and the mass of dense gas, though this interpretation is subject to several caveats. A similar result was seen by Gao and Solomon (2004) and Wu et al. (2010) using HCN (1-0), a tracer with similar conditions for excitation as HCO$^+$ (3-2). Those results were tested using the Bayesian linear regression routine. Considering all sources with L$_{IR}$ exceeding 10$^3$ $\lsun$, the slope is 0.99 $\pm$ 0.01 (1.17 $\pm$ 0.11 for Galactic clumps and 1.07 $\pm$ 0.06 for galaxies). 

The results of our work are generally consistent with theoretical predictions from radiative transfer models. Two different models, one considering GMCs (Krumholz and Thompson 2007) and one considering isolated galaxies and galaxy mergers (Narayanan et al. 2008), predicted that the slope of the log(L$_{bol}$)-log(L$'$) relation should depend on the critical density of the tracer and the mean density of the gas. If the critical density is above the mean gas density, then only high density peaks are thermalized and emit. In this case, the line luminosity rises super-linearly with density and the log(L$_{bol}$)-log(L$'$) relation is linear. If the critical density of the tracer is below the mean gas density, then nearly all the gas is traced and the line luminosity faithfully traces the gas, rising linearly with density. This leads to a super-linear L$_{bol}$-L$'$ relation. Both models predict a roughly linear relation for HCN (1-0) and since HCO$^+$ (3-2) is excited in similar 
conditions, it should behave similarly.

The conversion of L$_{bol}$ to SFR has many caveats. It can be confounded by the presence of AGN boosting the luminosity by heating dust grains or radiatively pumping the gas tracers to cause excess line emission. Uncertainty in the age of the stellar populations and in the ability to sample the IMF plague Galactic clumps. Ultimately, high resolution mid-infrared and submillimeter observations are needed to resolve embedded infrared populations and dense gas cores in Galactic clumps to reliably determine the SFR.
All of these caveats must be considered and characterized before the underlying star formation relation can be fully understood.

Our study suffered from a small number of galaxies mapped in HCO$^+$ (3-2) as only 14 were available (Graci{\'a}-Carpio et al. 2008). A more complete survey must be conducted to extend the sparsely populated moderate luminosity region of the L$_{bol}$-L$'$ plot toward L$_{bol}$ $<$ 10$^{11}$ $\lsun$. A followup extragalatic survey should include galaxies that span a wider range in L$_{bol}$.

\acknowledgements{Acknowledgements}
The authors would like to acknowledge John Downey, Sean Keel, Robert Moulton for operating the HHT during our observations. We thank Brandon Kelly for explaining the details of the Bayesian linear regression program. We also thank Timothy Ellsworth-Bowers for sharing his
work in distance determination prior to publication. 
Yancy Shirley is partially supported by NSF grant AST-1008577.



\begin{deluxetable}{cccccc}
\tablecolumns{6}
\tabletypesize{\footnotesize}
\tablecaption{Source Properties\label{tab1}}
\tablewidth{0pt} 
\tablehead{
\colhead{Source}   & 
\colhead{$\alpha$ (J2000)} &
\colhead{$\delta$ (J2000)} &
\colhead{log({\lbol} or L$_{IR}$)}    &
\colhead{Distance} &
\colhead{Distance \footnotemark[1]} \\
\colhead{}   & 
\colhead{[h m s]} &
\colhead{[$\degree$ $\am$ $\as$]} &
\colhead{[$\lsun$]}    &
\colhead{[kpc]} &
\colhead{Comment}
}
\startdata 
019.884-053	&	18	29	14.3	&	-11	50	22.8	&	$	5.15	\pm	0.01	$	&	$	12.47	\pm	0.27	$	&	Complex	\\
028.861+006	&	18	43	46.3	&	-3	35	31.6	&	$	5.01	\pm	0.00	$	&	$	5.46	\pm	0.26	$	&	Complex	\\
027.187-008	&	18	41	13.2	&	-5	8	58.1	&	$	5.38	\pm	0.01	$	&	$	12.93	\pm	0.33	$	&	Complex	\\
029.436-017	&	18	45	40.6	&	-3	11	21.4	&	$	3.44	\pm	0.03	$	&	$	4.87	\pm	0.24	$	&	IRDC	\\
019.923-025	&	18	28	18.9	&	-11	40	31.2	&	$	3.48	\pm	0.02	$	&	$	4.25	\pm	0.21	$	&	IRDC-Complex	\\
024.730+015	&	18	35	50.6	&	-7	13	26.4	&	$	4.51	\pm	0.02	$	&	$	5.77	\pm	0.20	$	&	Complex	\\
034.712-059	&	18	56	48.2	&	1	18	45.6	&	$	3.83	\pm	0.01	$	&	$	2.80	\pm	0.28	$	&	Complex	\\
027.925+020	&	18	41	34.5	&	-4	21	7.9	&	$	3.83	\pm	0.01	$	&	$	2.86	\pm	0.29	$	&	Complex	\\
037.555+019	&	18	59	10	&	4	12	18.4	&	$	4.64	\pm	0.01	$	&	$	5.59	\pm	0.37	$	&	Complex	\\
059.497-023	&	19	43	42.3	&	23	20	19	&	$	3.48	\pm	0.01	$	&	$	3.27	\pm	0.68	$	&	IRDC	\\
033.384+000	&	18	52	14.5	&	0	24	54.5	&	$	4.20	\pm	0.07	$	&	$	6.35	\pm	0.34	$	&	IRDC	\\
025.803-015	&	18	38	56.5	&	-6	24	53.4	&	$	4.71	\pm	0.01	$	&	$	5.20	\pm	0.22	$	&	Complex	\\
025.411+010	&	18	37	17	&	-6	38	28.1	&	$	3.97	\pm	0.02	$	&	$	5.26	\pm	0.20	$	&	Complex	\\
059.355-020	&	19	43	17.9	&	23	13	58.9	&	$	4.11	\pm	0.01	$	&	$	4.04	\pm	0.68	$	&	Complex	\\
025.393+004	&	18	37	30.3	&	-6	41	14.8	&	$	3.37	\pm	0.01	$	&	$	1.11	\pm	0.28	$	&		\\
023.385+018	&	18	33	14.6	&	-8	23	55.6	&	$	4.33	\pm	0.01	$	&	$	4.51	\pm	0.21	$	&	Complex	\\
\tableline																							
121.30+0.66	&	0	36	48	&	63	29	1	&	$	3.09	\pm	0.02	$	&	$	0.85	\pm	0.42	$	&		\\
123.07-6.31	&	0	52	25	&	56	33	53	&	$	3.74	\pm	0.04	$	&	$	2.2	\pm	0.4	$	&		\\
W3(OH)	&	2	27	5	&	61	52	26	&	$	4.99	\pm	0.01	$	&	$	2.4	\pm	0.4	$	&		\\
S231	&	5	39	13	&	35	45	54	&	$	4.00	\pm	0.02	$	&	$	2.0	\pm	0.4	$	&		\\
S252A	&	6	8	35	&	20	39	3	&	$	4.32	\pm	0.00	$	&	$	1.5	\pm	0.4	$	&		\\
RCW142	&	17	50	15	&	-28	54	32	&	$	4.68	\pm	0.04	$	&	$	2.0	\pm	0.4	$	&		\\
W28A2	&	18	0	30	&	-24	3	58	&	$	5.44	\pm	0.05	$	&	$	2.6	\pm	0.4	$	&		\\
M8E	&	18	4	53	&	-24	26	42	&	$	4.21	\pm	0.02	$	&	$	1.8	\pm	0.4	$	&		\\
8.67-0.36	&	18	6	19	&	-21	37	38	&	$	5.08	\pm	0.03	$	&	$	4.5	\pm	0.5	$	&		\\
10.60-0.40	&	18	10	29	&	-19	55	49	&	$	5.99	\pm	0.04	$	&	$	6.5	\pm	0.5	$	&		\\
12.89+0.49	&	18	11	51	&	-17	31	31	&	$	4.55	\pm	0.01	$	&	$	3.5	\pm	0.4	$	&		\\
W33A	&	18	14	39	&	-17	52	11	&	$	5.12	\pm	0.01	$	&	$	4.5	\pm	0.4	$	&		\\
24.49-0.04	&	18	36	5	&	-7	31	23	&	$	4.69	\pm	0.01	$	&	$	3.5	\pm	0.2	$	&		\\
W43S	&	18	46	4	&	-2	39	26	&	$	6.10	\pm	0.02	$	&	$	8.5	\pm	0.3	$	&		\\
31.41+0.31	&	18	47	34	&	-1	12	46	&	$	5.27	\pm	0.03	$	&	$	7.9	\pm	0.3	$	&		\\
W44	&	18	53	18	&	1	14	57	&	$	5.58	\pm	0.03	$	&	$	3.7	\pm	0.3	$	&		\\
40.50+2.54	&	18	56	10	&	7	53	14	&	$	4.29	\pm	0.05	$	&	$	2.1	\pm	0.4	$	&		\\
35.20-0.74	&	18	58	13	&	1	40	36	&	$	4.76	\pm	0.05	$	&	$	3.3	\pm	0.4	$	&		\\
59.78+0.06	&	19	43	12	&	23	43	54	&	$	4.02	\pm	0.03	$	&	$	2.2	\pm	0.6	$	&		\\
ON1	&	20	10	9	&	31	31	37	&	$	5.11	\pm	0.02	$	&	$	6.0	\pm	0.4	$	&		\\
ON2S	&	20	21	41	&	37	25	29	&	$	5.59	\pm	0.02	$	&	$	5.5	\pm	0.4	$	&		\\
W75N	&	20	38	37	&	42	37	37	&	$	5.29	\pm	0.03	$	&	$	3.0	\pm	0.4	$	&		\\
W75OH	&	20	39	1	&	42	22	50	&	$	4.40	\pm	0.02	$	&	$	3.0	\pm	0.4	$	&		\\
S140	&	22	19	18	&	63	18	47	&	$	4.34	\pm	0.02	$	&	$	0.9	\pm	0.4	$	&		\\
CepA	&	22	56	18	&	62	1	46	&	$	4.52	\pm	0.02	$	&	$	0.73	\pm	0.42	$	&		\\
NGC7538-IRS9	&	23	14	2	&	61	27	20	&	$	4.59	\pm	0.05	$	&	$	2.8	\pm	0.7	$	&		\\
S157	&	23	16	4	&	60	1	41	&	$	4.31	\pm	0.03	$	&	$	2.5	\pm	0.7	$	&		\\
\tableline
Mean		&			&				&		5.09				&		4.11				&		\\
Standard Deviation	&		&				&		0.87				&		2.67				&		\\
Median		&			&				&		4.52				&		3.5				&	
\enddata

{$^1$IRDC: The source is associated with an infrared dark cloud. The near kinematic distance was assigned. Complex: The source is a part of a complex with a known distance. If one of the kinematic distances was similar to the complex distance, that kinematic distance was assigned (Mottram et al. 2010).}
\end{deluxetable}

\begin{deluxetable}{cccccc}
\tablecolumns{6}
\tabletypesize{\footnotesize}
\tablecaption{$HCO^{+}$ Properties \label{tab2}}
\tablewidth{0pt} 
\tablehead{
\colhead{Source} &
\colhead{I} &
\colhead{$v_{LSR}$} &
\colhead{$\theta_{s*b}$} &
\colhead{R} &
\colhead{log(L$'$)} \\
\colhead{} &
\colhead{[$K$ $km$ $s^{-1}$]} &
\colhead{[$km$ $s^{-1}$]} &
\colhead{[$\as$]} &
\colhead{[pc]} &
\colhead{[$K$ $km$ $s^{-1}$ $pc^{-2}$]}
}
\startdata 
019.884-053	&	$	46.12	\pm	2.47	$	&	43.99	&	$	41.8	\pm	3.3	$	&	$	0.96	\pm	0.13	$	&	$	2.52	\pm	0.07	$	\\
028.861+006	&	$	16.91	\pm	0.94	$	&	103.90	&	$	38.6	\pm	3.1	$	&	$	0.36	\pm	0.06	$	&	$	1.30	\pm	0.09	$	\\
027.187-008	&	$	15.07	\pm	0.87	$	&	25.65	&	$	29.3	\pm	3.1	$	&	$	0.34	\pm	0.27	$	&	$	1.76	\pm	0.10	$	\\
029.436-017	&	$	3.17	\pm	0.36	$	&	86.19	&	$	36.7	\pm	11.0	$	&	$	0.29	\pm	0.19	$	&	$	0.43	\pm	0.27	$	\\
019.923-025	&	$	7.54	\pm	0.48	$	&	64.30	&	$	46.3	\pm	7.8	$	&	$	0.39	\pm	0.10	$	&	$	0.89	\pm	0.16	$	\\
024.730+015	&	$	5.65	\pm	0.44	$	&	109.61	&	$	32.6	\pm	6.1	$	&	$	0.25	\pm	0.16	$	&	$	0.73	\pm	0.17	$	\\
034.712-059	&	$	9.03	\pm	0.59	$	&	44.17	&	$	46.4	\pm	5.5	$	&	$	0.26	\pm	0.05	$	&	$	0.61	\pm	0.14	$	\\
027.925+020	&	$	12.45	\pm	0.76	$	&	42.65	&	$	34.8	\pm	2.7	$	&	$	0.15	\pm	0.03	$	&	$	0.52	\pm	0.11	$	\\
037.555+019	&	$	13.67	\pm	0.83	$	&	86.21	&	$	44.0	\pm	4.5	$	&	$	0.47	\pm	0.08	$	&	$	1.34	\pm	0.11	$	\\
059.497-023	&	$	17.03	\pm	0.97	$	&	27.13	&	$	44.4	\pm	5.1	$	&	$	0.28	\pm	0.08	$	&	$	0.98	\pm	0.21	$	\\
033.384+000	&	$	11.14	\pm	0.67	$	&	103.88	&	$	40.5	\pm	4.2	$	&	$	0.46	\pm	0.09	$	&	$	1.29	\pm	0.10	$	\\
025.803-015	&	$	16.41	\pm	0.94	$	&	92.06	&	$	39.4	\pm	3.4	$	&	$	0.36	\pm	0.06	$	&	$	1.26	\pm	0.09	$	\\
025.411+010	&	$	12.10	\pm	0.74	$	&	95.91	&	$	37.7	\pm	3.2	$	&	$	0.33	\pm	0.06	$	&	$	1.10	\pm	0.09	$	\\
059.355-020	&	$	10.52	\pm	0.69	$	&	29.53	&	$	49.0	\pm	5.0	$	&	$	0.40	\pm	0.09	$	&	$	1.04	\pm	0.17	$	\\
025.393+004	&	$	11.27	\pm	0.66	$	&	-13.02	&	$	34.9	\pm	2.8	$	&	$	0.06	\pm	0.02	$	&	$	-0.34	\pm	0.23	$	\\
023.385+018	&	$	9.96	\pm	0.71	$	&	75.75	&	$	37.9	\pm	4.0	$	&	$	0.29	\pm	0.06	$	&	$	0.89	\pm	0.11	$	\\
\tableline
Mean		&		13.63				&		&						&		0.35				&		1.53				\\
Standard Deviation	&	9.51				&		&						&		0.19				&		1.05				\\
Median		&		11.69				&		&						&		0.34				&		1.01
\enddata
\end{deluxetable}

\begin{deluxetable}{cccc}
\tablecolumns{4}
\tabletypesize{\footnotesize}
\tablecaption{$HCO^{+}$ Properties \label{tab3}}
\tablewidth{0pt} 
\tablehead{
\colhead{Source} &
\colhead{$\Delta v$} &
\colhead{log(M$_{iso}$)} &
\colhead{log(M$_{vir}$)} \\
\colhead{} &
\colhead{[$km$ $s^{-1}$]} &
\colhead{[$\msun$]} &
\colhead{[$\msun$]}
}
\startdata
019.884-053	&	$	5.8	\pm	0.1	$	&	$	3.99	\pm	0.04	$	&	$	3.69	\pm	0.06	$	\\
028.861+006	&	$	5.7	\pm	0.1	$	&	$	3.05	\pm	0.05	$	&	$	3.25	\pm	0.07	$	\\
027.187-008	&	$	7.4	\pm	0.3	$	&	$	3.63	\pm	0.05	$	&	$	3.44	\pm	0.35	$	\\
029.436-017	&	$	3.4	\pm	0.2	$	&	$	2.28	\pm	0.08	$	&	$	2.70	\pm	0.30	$	\\
019.923-025	&	$	4.8	\pm	0.2	$	&	$	2.71	\pm	0.05	$	&	$	3.13	\pm	0.12	$	\\
024.730+015	&	$	3.7	\pm	0.2	$	&	$	2.61	\pm	0.07	$	&	$	2.72	\pm	0.27	$	\\
034.712-059	&	$	9.0	\pm	0.7	$	&	$	2.20	\pm	0.09	$	&	$	3.50	\pm	0.11	$	\\
027.925+020	&	$	4.9	\pm	0.2	$	&	$	2.06	\pm	0.10	$	&	$	2.73	\pm	0.10	$	\\
037.555+019	&	$	8.9	\pm	0.3	$	&	$	2.98	\pm	0.07	$	&	$	3.75	\pm	0.08	$	\\
059.497-023	&	$	3.4	\pm	0.1	$	&	$	2.10	\pm	0.20	$	&	$	2.69	\pm	0.12	$	\\
033.384+000	&	$	4.8	\pm	0.2	$	&	$	2.99	\pm	0.06	$	&	$	3.20	\pm	0.09	$	\\
025.803-015	&	$	3.7	\pm	0.2	$	&	$	2.67	\pm	0.06	$	&	$	2.86	\pm	0.09	$	\\
025.411+010	&	$	6.9	\pm	0.2	$	&	$	2.52	\pm	0.07	$	&	$	3.38	\pm	0.08	$	\\
059.355-020	&	$	2.8	\pm	0.1	$	&	$	1.82	\pm	0.26	$	&	$	2.67	\pm	0.10	$	\\
025.393+004	&	$	7.3	\pm	0.2	$	&	$	1.37	\pm	0.22	$	&	$	2.68	\pm	0.14	$	\\
023.385+018	&	$	3.8	\pm	0.1	$	&	$	2.58	\pm	0.06	$	&	$	2.80	\pm	0.10	$	\\

\tableline
Mean		&						&		3.09				&		3.24				\\
Standard Deviation	&					&		0.87				&		0.41				\\
Median		&						&		2.60				&		3.01
\enddata
\end{deluxetable}

\begin{deluxetable}{ccccc}
\tablecolumns{5}
\tabletypesize{\footnotesize}
\tablecaption{Photometry \label{tab4}}
\tablewidth{0pt} 
\tablehead{
\colhead{Source} &
\colhead{$\lambda$} &
\colhead{Flux} &
\colhead{Beam} &
\colhead{Ref \footnotemark[1]} \\
\colhead{} &
\colhead{[$\mu$m]} &
\colhead{[Jy]} &
\colhead{[\arcsec]} &
\colhead{} 
}
\startdata
019.884-053	&	8	&	$	0.61	\pm	0.03	$	&	18	&	1	\\
	&	14	&	$	0.73	\pm	0.06	$	&	18	&	1	\\
	&	21	&	$	3.43	\pm	0.22	$	&	18	&	1	\\
	&	25	&		11.07				&	300x45	&	2	\\
	&	60	&	$	394.45	\pm	3.75	$	&	165x87	&	3	\\
	&	65	&	$	285.49	\pm	41.20	$	&	600x60	&	4	\\
	&	70	&	$	407.22	\pm	2.99	$	&	62.4	&	3	\\
	&	90	&	$	211.59	\pm	38.44	$	&	600x90	&	4	\\
	&	100	&	$	1155.65	\pm	6.07	$	&	195x150	&	3	\\
	&	140	&	$	598.96	\pm	37.25	$	&	700x100	&	4	\\
	&	160	&	$	795	\pm	115.70	$	&	700x150	&	4	\\
	&	850	&		8.24				&	13.5	&	5	\\
	&	1100	&	$	4.91	\pm	0.37	$	&	33	&	6	\\
028.861+006	&	8	&	$	3.28	\pm	0.13	$	&	18	&	1	\\
	&	12	&	$	7.12	\pm	0.36	$	&	18	&	1	\\
	&	14	&	$	15.34	\pm	0.94	$	&	18	&	1	\\
	&	21	&	$	34.01	\pm	2.04	$	&	18	&	1	\\
	&	25	&		105.20				&	300x45	&	2	\\
	&	60	&	$	1945.55	\pm	7.08	$	&	165x87	&	3	\\
	&	65	&	$	1493.04	\pm	176.25	$	&	600x60	&	4	\\
	&	70	&	$	1567.44	\pm	3.73	$	&	62.4	&	3	\\
	&	90	&	$	1207.87	\pm	88.65	$	&	600x90	&	4	\\
	&	100	&	$	3263.03	\pm	19.92	$	&	195x150	&	3	\\
	&	140	&	$	839.89	\pm	2.29	$	&	700x100	&	4	\\
	&	1100	&	$	4.39	\pm	0.29	$	&	33	&	6	\\
027.187-008	&	8	&	$	1.35	\pm	0.06	$	&	18	&	1	\\
	&	12	&	$	3.06	\pm	0.16	$	&	18	&	1	\\
	&	14	&	$	5.22	\pm	0.32	$	&	18	&	1	\\
	&	21	&	$	16.77	\pm	1.01	$	&	18	&	1	\\
	&	25	&		39.20				&	300x45	&	2	\\
	&	60	&	$	797.61	\pm	6.92	$	&	165x87	&	3	\\
	&	65	&	$	527.15	\pm	98.70	$	&	600x60	&	4	\\
	&	70	&	$	640.54	\pm	3.38	$	&	62.4	&	3	\\
	&	90	&	$	360.93	\pm	120.56	$	&	600x90	&	4	\\
	&	100	&	$	1634.41	\pm	14.90	$	&	195x150	&	3	\\
	&	140	&	$	447.09	\pm	32.62	$	&	700x100	&	4	\\
	&	160	&	$	562.54	\pm	84.36	$	&	700x150	&	4	\\
	&	1100	&	$	2.20	\pm	0.22	$	&	33	&	6	\\
029.436-017	&	8	&	$	1.33	\pm	0.05	$	&	18	&	1	\\
	&	12	&	$	1.60	\pm	0.10	$	&	18	&	1	\\
	&	14	&	$	2.33	\pm	0.15	$	&	18	&	1	\\
	&	21	&	$	3.18	\pm	0.20	$	&	18	&	1	\\
	&	65	&	$	22.21	\pm	3.15	$	&	600x60	&	4	\\
	&	70	&	$	18.39	\pm	4.16	$	&	62.4	&	3	\\
	&	90	&	$	24.28	\pm	0.65	$	&	600x90	&	4	\\
	&	160	&	$	130.82	\pm	22.81	$	&	700x150	&	4	\\
	&	1100	&	$	0.67	\pm	0.14	$	&	33	&	6	\\
019.923-025	&	8	&	$	0.95	\pm	0.04	$	&	18	&	1	\\
	&	12	&	$	2.33	\pm	0.13	$	&	18	&	1	\\
	&	14	&	$	4.15	\pm	0.25	$	&	18	&	1	\\
	&	21	&	$	5.50	\pm	0.34	$	&	18	&	1	\\
	&	25	&		6.74				&	300x45	&	2	\\
	&	60	&	$	13.94	\pm	2.27	$	&	165x87	&	3	\\
	&	70	&	$	35.40	\pm	3.19	$	&	62.4	&	3	\\
	&	90	&	$	32.14	\pm	1.06	$	&	600x90	&	4	\\
	&	100	&	$	315.19	\pm	23.91	$	&	195x150	&	3	\\
	&	140	&	$	96.56	\pm	9.75	$	&	700x100	&	4	\\
	&	160	&	$	140.93	\pm	22.97	$	&	700x150	&	4	\\
	&	1100	&	$	3.02	\pm	0.22	$	&	33	&	6	\\
024.730+015	&	8	&	$	1.17	\pm	0.05	$	&	18	&	1	\\
	&	12	&	$	1.52	\pm	0.08	$	&	18	&	1	\\
	&	14	&	$	1.74	\pm	0.11	$	&	18	&	1	\\
	&	21	&	$	5.44	\pm	0.33	$	&	18	&	1	\\
	&	25	&		15.20				&	300x45	&	2	\\
	&	60	&		325.50				&	90x300	&	2	\\
	&	65	&	$	177.78	\pm	7.83	$	&	600x60	&	4	\\
	&	70	&	$	188.39	\pm	4.32	$	&	62.4	&	3	\\
	&	100	&		1279				&	180x300	&	2	\\
	&	140	&	$	649.83	\pm	62.51	$	&	700x100	&	4	\\
	&	160	&	$	749.97	\pm	72.18	$	&	700x150	&	4	\\
	&	1100	&	$	1.39	\pm	0.19	$	&	33	&	6	\\
034.712-059	&	8	&	$	8.13	\pm	0.33	$	&	18	&	1	\\
	&	12	&	$	10.26	\pm	0.51	$	&	18	&	1	\\
	&	14	&	$	13.00	\pm	0.79	$	&	18	&	1	\\
	&	21	&	$	26.89	\pm	1.61	$	&	18	&	1	\\
	&	25	&		43.70				&	300x45	&	2	\\
	&	60	&	$	290.98	\pm	2.48	$	&	165x87	&	3	\\
	&	65	&	$	295.41	\pm	63.67	$	&	600x60	&	4	\\
	&	70	&	$	250.61	\pm	3.20	$	&	62.4	&	3	\\
	&	90	&	$	178.62	\pm	65.62	$	&	600x90	&	4	\\
	&	100	&	$	1086.15	\pm	9.72	$	&	195x150	&	3	\\
	&	140	&	$	265.17	\pm	33.81	$	&	700x100	&	4	\\
	&	1100	&		2.46				&	33	&	6	\\
027.925+020	&	8	&	$	0.94	\pm	0.04	$	&	18	&	1	\\
	&	12	&	$	1.14	\pm	0.08	$	&	18	&	1	\\
	&	14	&	$	1.68	\pm	0.11	$	&	18	&	1	\\
	&	21	&	$	6.49	\pm	0.40	$	&	18	&	1	\\
	&	60	&	$	402.72	\pm	2.33	$	&	165x87	&	3	\\
	&	65	&	$	264.64	\pm	22.53	$	&	600x60	&	4	\\
	&	70	&	$	218.93	\pm	3.17	$	&	62.4	&	3	\\
	&	90	&	$	275.78	\pm	2.76	$	&	600x90	&	4	\\
	&	100	&	$	1137.67	\pm	10.42	$	&	195x150	&	3	\\
	&	140	&	$	304.94	\pm	53.19	$	&	700x100	&	4	\\
	&	160	&	$	404.62	\pm	31.68	$	&	700x150	&	4	\\
	&	1100	&	$	1.86	\pm	0.17	$	&	33	&	6	\\
037.555+019	&	8	&	$	8.57	\pm	0.35	$	&	18	&	1	\\
	&	12	&	$	14.83	\pm	0.74	$	&	18	&	1	\\
	&	14	&	$	24.74	\pm	1.51	$	&	18	&	1	\\
	&	21	&	$	37.87	\pm	2.27	$	&	18	&	1	\\
	&	25	&		85.25				&	300x45	&	2	\\
	&	60	&	$	538.36	\pm	5.57	$	&	165x87	&	3	\\
	&	65	&	$	380.24	\pm	128.67	$	&	600x60	&	4	\\
	&	70	&	$	510.17	\pm	3.26	$	&	62.4	&	3	\\
	&	90	&	$	288.24	\pm	104.53	$	&	600x90	&	4	\\
	&	100	&	$	937.88	\pm	10.59	$	&	195x150	&	3	\\
	&	140	&	$	429.70	\pm	86.90	$	&	700x100	&	4	\\
	&	160	&	$	537.72	\pm	143.92	$	&	700x150	&	4	\\
	&	850	&		5.45				&	13.5	&	5	\\
	&	1100	&	$	3.11	\pm	0.25	$	&	33	&	6	\\
059.497-023	&	8	&	$	0.08	\pm	0.01	$	&	18	&	1	\\
	&	12	&	$	1.07	\pm	0.52	$	&	18	&	1	\\
	&	14	&	$	1.65	\pm	0.11	$	&	18	&	1	\\
	&	21	&	$	4.66	\pm	0.29	$	&	18	&	1	\\
	&	25	&		10.71				&	300x45	&	2	\\
	&	60	&	$	122.40	\pm	1.07	$	&	165x87	&	3	\\
	&	65	&	$	64.53	\pm	9.10	$	&	600x60	&	4	\\
	&	70	&	$	33.92	\pm	3.98	$	&	62.4	&	3	\\
	&	90	&	$	37.47	\pm	7.78	$	&	600x90	&	4	\\
	&	100	&	$	462.16	\pm	4.80	$	&	195x150	&	3	\\
	&	140	&	$	96.32	\pm	39.96	$	&	700x100	&	4	\\
	&	160	&	$	192.87	\pm	20.75	$	&	700x150	&	4	\\
	&	1100	&	$	1.37	\pm	0.23	$	&	33	&	6	\\
033.384+000	&	8	&	$	0.07	\pm	0.01	$	&	18	&	1	\\
	&	12	&	$	0.21	\pm	0.02	$	&	18	&	1	\\
	&	14	&	$	1.22	\pm	0.07	$	&	18	&	1	\\
	&	21	&	$	5.27	\pm	0.32	$	&	18	&	1	\\
	&	65	&	$	122.77	\pm	35.07	$	&	600x60	&	4	\\
	&	70	&	$	124.36	\pm	4.99	$	&	62.4	&	3	\\
	&	90	&	$	145.52	\pm	15.97	$	&	600x90	&	4	\\
	&	140	&	$	279.37	\pm	76.98	$	&	700x100	&	4	\\
	&	160	&	$	399.49	\pm	143.59	$	&	700x150	&	4	\\
	&	1100	&	$	4.13	\pm	0.28	$	&	33	&	6	\\
025.803-015	&	8	&	$	2.26	\pm	0.09	$	&	18	&	1	\\
	&	12	&	$	8.55	\pm	0.43	$	&	18	&	1	\\
	&	14	&	$	18.27	\pm	1.11	$	&	18	&	1	\\
	&	21	&	$	68.66	\pm	4.12	$	&	18	&	1	\\
	&	60	&	$	790.50	\pm	6.81	$	&	165x87	&	3	\\
	&	65	&	$	349.18	\pm	35.48	$	&	600x60	&	4	\\
	&	70	&	$	417.27	\pm	4.03	$	&	62.4	&	3	\\
	&	90	&	$	345.45	\pm	60.75	$	&	600x90	&	4	\\
	&	100	&	$	1846.48	\pm	16.71	$	&	195x150	&	3	\\
	&	140	&	$	351.46	\pm	120.18	$	&	700x100	&	4	\\
	&	160	&	$	282.21	\pm	135.45	$	&	700x150	&	4	\\
	&	850	&		4.95				&	13.5	&	5	\\
	&	1100	&	$	1.90	\pm	0.19	$	&	33	&	6	\\
025.411+010	&	8	&	$	2.33	\pm	0.10	$	&	18	&	1	\\
	&	12	&	$	3.46	\pm	0.17	$	&	18	&	1	\\
	&	14	&	$	4.70	\pm	0.29	$	&	18	&	1	\\
	&	21	&	$	8.43	\pm	0.51	$	&	18	&	1	\\
	&	25	&		17.50				&	300x45	&	2	\\
	&	60	&		109.90				&	90x300	&	2	\\
	&	65	&	$	152.96	\pm	10.85	$	&	600x60	&	4	\\
	&	70	&	$	142.10	\pm	3.36	$	&	62.4	&	3	\\
	&	90	&	$	75.81	\pm	11.84	$	&	600x90	&	4	\\
	&	140	&	$	157.19	\pm	43.69	$	&	700x100	&	4	\\
	&	160	&	$	279.51	\pm	30.24	$	&	700x150	&	4	\\
	&	1100	&	$	1.01	\pm	0.16	$	&	33	&	6	\\
059.355-020	&	8	&	$	1.82	\pm	0.07	$	&	18	&	1	\\
	&	12	&	$	5.24	\pm	0.27	$	&	18	&	1	\\
	&	14	&	$	8.72	\pm	0.53	$	&	18	&	1	\\
	&	21	&	$	25.60	\pm	1.54	$	&	18	&	1	\\
	&	25	&		41.16				&	300x45	&	2	\\
	&	60	&	$	333.15	\pm	1.90	$	&	165x87	&	3	\\
	&	70	&	$	87.50	\pm	3.36	$	&	62.4	&	3	\\
	&	100	&	$	879.54	\pm	7.21	$	&	195x150	&	3	\\
	&	140	&	$	187.37	\pm	40.37	$	&	700x100	&	4	\\
	&	1100	&	$	0.38	\pm	0.20	$	&	33	&	6	\\
025.393+004	&	8	&	$	2.29	\pm	0.09	$	&	18	&	1	\\
	&	12	&	$	4.78	\pm	0.24	$	&	18	&	1	\\
	&	14	&	$	7.14	\pm	0.44	$	&	18	&	1	\\
	&	21	&	$	17.00	\pm	1.02	$	&	18	&	1	\\
	&	60	&	$	1116.91	\pm	8.23	$	&	165x87	&	3	\\
	&	65	&	$	430.67	\pm	115.07	$	&	600x60	&	4	\\
	&	70	&	$	804.15	\pm	4.61	$	&	62.4	&	3	\\
	&	90	&	$	307.23	\pm	59.00	$	&	600x90	&	4	\\
	&	100	&	$	2350.42	\pm	26.44	$	&	195x150	&	3	\\
	&	140	&	$	569.92	\pm	45.48	$	&	700x100	&	4	\\
	&	160	&	$	663.00	\pm	90.06	$	&	700x150	&	4	\\
	&	1100	&	$	2.31	\pm	0.22	$	&	33	&	6	\\
023.385+018	&	8	&	$	23.47	\pm	0.96	$	&	18	&	1	\\
	&	12	&	$	39.00	\pm	1.95	$	&	18	&	1	\\
	&	14	&	$	50.99	\pm	3.11	$	&	18	&	1	\\
	&	21	&	$	54.71	\pm	3.28	$	&	18	&	1	\\
	&	25	&		76.45				&	300x45	&	2	\\
	&	60	&	$	281.63	\pm	2.84	$	&	165x87	&	3	\\
	&	65	&	$	262.22	\pm	38.11	$	&	600x60	&	4	\\
	&	70	&	$	239.14	\pm	3.44	$	&	62.4	&	3	\\
	&	100	&	$	157.45	\pm	16.07	$	&	195x150	&	3	\\
	&	160	&	$	331.91	\pm	25.58	$	&	700x150	&	4	\\
	&	1100	&	$	2.26	\pm	0.21	$	&	33	&	6	\\
\tableline													
121.30+0.66	&	12	&	$	1.8	\pm	0.20	$	&	300x45	&	2	\\
	&	25	&	$	21	\pm	1.00	$	&	300x45	&	2	\\
	&	60	&	$	357	\pm	21.00	$	&	90x300	&	2	\\
	&	90	&	$	790.33	\pm	88.03	$	&	600x90	&	4	\\
	&	100	&	$	685	\pm	55.00	$	&	180x300	&	2	\\
	&	140	&	$	843.33	\pm	70.25	$	&	700x100	&	4	\\
	&	143	&	$	1615	\pm	323.00	$	&	105	&	7	\\
	&	185	&	$	2317	\pm	463.00	$	&	102	&	7	\\
	&	350	&	$	410	\pm	82.00	$	&	120	&	8	\\
	&	450	&	$	49	\pm	2.00	$	&	18	&	9	\\
	&	800	&	$	6.2	\pm	0.02	$	&	16	&	9	\\
	&	850	&	$	17	\pm	3.40	$	&	18	&	10	\\
	&	1100	&	$	5.6	\pm	1.10	$	&	18	&	11	\\
123.07-6.31	&	25	&	$	13	\pm	1.00	$	&	300x45	&	2	\\
	&	60	&	$	330	\pm	46.00	$	&	90x300	&	2	\\
	&	65	&	$	263.84	\pm	73.24	$	&	600x60	&	4	\\
	&	100	&	$	1166	\pm	117.00	$	&	180x300	&	2	\\
	&	140	&	$	796.87	\pm	277.15	$	&	700x100	&	4	\\
	&	160	&	$	784.67	\pm	319.67	$	&	700x150	&	4	\\
	&	350	&	$	290	\pm	58.00	$	&	120	&	8	\\
W3(OH)	&	12	&	$	40.58	\pm	4.06	$	&	300x45	&	2	\\
	&	20	&	$	270	\pm	27.00	$	&	49	&	8	\\
	&	25	&	$	670	\pm	67.00	$	&	49	&	8	\\
	&	30	&	$	1400	\pm	140.00	$	&	49	&	8	\\
	&	40	&	$	4000	\pm	400.00	$	&	49	&	8	\\
	&	60	&	$	7000	\pm	700.00	$	&	49	&	8	\\
	&	65	&	$	6918.19	\pm	198.06	$	&	600x60	&	4	\\
	&	70	&	$	8500	\pm	850.00	$	&	49	&	8	\\
	&	100	&	$	9000	\pm	900.00	$	&	49	&	8	\\
	&	140	&	$	6700	\pm	670.00	$	&	49	&	8	\\
	&	160	&	$	8645.55	\pm	1006.47	$	&	700x150	&	4	\\
	&	180	&	$	4900	\pm	490.00	$	&	49	&	8	\\
	&	200	&	$	4100	\pm	410.00	$	&	49	&	8	\\
	&	250	&	$	2400	\pm	240.00	$	&	49	&	8	\\
	&	300	&	$	1400	\pm	140.00	$	&	49	&	8	\\
	&	350	&	$	1130	\pm	230.00	$	&	120	&	8	\\
	&	500	&	$	250	\pm	25.00	$	&	49	&	8	\\
	&	600	&	$	135	\pm	14.00	$	&	49	&	8	\\
	&	800	&	$	51	\pm	5.10	$	&	49	&	8	\\
	&	1000	&	$	24	\pm	2.40	$	&	49	&	8	\\
S231	&	12	&	$	5.6	\pm	0.20	$	&	300x45	&	2	\\
	&	25	&	$	75	\pm	3.70	$	&	300x45	&	2	\\
	&	60	&	$	722	\pm	72.00	$	&	90x300	&	2	\\
	&	100	&	$	1310	\pm	131.00	$	&	180x300	&	2	\\
	&	160	&	$	1841.75	\pm	94.64	$	&	700x150	&	4	\\
	&	350	&	$	522	\pm	100.00	$	&	120	&	8	\\
S252A	&	12	&	$	16	\pm	0.60	$	&	300x45	&	2	\\
	&	25	&	$	77	\pm	3.00	$	&	300x45	&	2	\\
	&	60	&	$	10321	\pm	34.00	$	&	90x300	&	2	\\
	&	100	&	$	1715	\pm	189.00	$	&	180x300	&	2	\\
	&	140	&	$	844.94	\pm	131.43	$	&	700x100	&	4	\\
	&	350	&	$	320	\pm	64.00	$	&	120	&	8	\\
RCW142	&	12	&	$	<	42		$	&	300x45	&	2	\\
	&	25	&	$	<	281		$	&	300x45	&	2	\\
	&	60	&	$	5476	\pm	986.00	$	&	90x300	&	2	\\
	&	65	&	$	2623.98	\pm	59.43	$	&	600x60	&	4	\\
	&	90	&	$	1616.73	\pm	134.92	$	&	600x90	&	4	\\
	&	100	&	$	13129	\pm	1313.00	$	&	180x300	&	2	\\
	&	160	&	$	8480.46	\pm	824.94	$	&	700x150	&	4	\\
	&	350	&	$	670	\pm	130.00	$	&	120	&	8	\\
W28A2	&	12	&	$	199	\pm	12.00	$	&	300x45	&	2	\\
	&	25	&	$	2190	\pm	131.00	$	&	300x45	&	2	\\
	&	60	&	$	12790	\pm	3198.00	$	&	90x300	&	2	\\
	&	100	&	$	26780	\pm	6695.00	$	&	180x300	&	2	\\
	&	350	&	$	1580	\pm	320.00	$	&	120	&	8	\\
M8E	&	12	&	$	118.6	\pm	11.86	$	&	300x45	&	2	\\
	&	25	&	$	289	\pm	17.00	$	&	300x45	&	2	\\
	&	60	&	$	1611	\pm	226.00	$	&	90x300	&	2	\\
	&	65	&	$	658.41	\pm	610.01	$	&	600x60	&	4	\\
	&	69	&		2600				&	54	&	12	\\
	&	90	&	$	805.25	\pm	355.31	$	&	600x90	&	4	\\
	&	100	&	$	2783	\pm	696.00	$	&	180x300	&	2	\\
	&	140	&	$	1570.57	\pm	90.78	$	&	700x100	&	4	\\
	&	160	&	$	2481.10	\pm	263.06	$	&	700x150	&	4	\\
	&	350	&	$	380	\pm	76.00	$	&	120	&	8	\\
	&	450	&	$	42	\pm	15.80	$	&	19	&	13	\\
	&	850	&	$	9	\pm	1.01	$	&	25	&	13	\\
  8.67-0.36	&	12	&	$	18.95	\pm	1.90	$	&	300x45	&	2	\\
	&	25	&	$	254	\pm	8.00	$	&	300x45	&	2	\\
	&	60	&	$	1895	\pm	303.00	$	&	90x300	&	2	\\
	&	65	&	$	1151.07	\pm	94.82	$	&	600x60	&	4	\\
	&	100	&	$	5125	\pm	1128.00	$	&	180x300	&	2	\\
	&	140	&	$	2623.88	\pm	312.83	$	&	700x100	&	4	\\
	&	350	&	$	650	\pm	130.00	$	&	120	&	8	\\
	&	450	&	$	390	\pm	98.00	$	&	18	&	10	\\
	&	850	&	$	49	\pm	10.00	$	&	18	&	10	\\
	&	1300	&	$	7.1	\pm	0.71	$	&	90	&	14	\\
10.60-0.40	&	12	&	$	<	23		$	&	300x45	&	2	\\
	&	25	&	$	<	148		$	&	300x45	&	2	\\
	&	60	&	$	9479	\pm	948.00	$	&	90x300	&	2	\\
	&	69	&		14000				&	90	&	15	\\
	&	100	&	$	21375	\pm	3847.00	$	&	180x300	&	2	\\
	&	350	&	$	1900	\pm	380.00	$	&	120	&	8	\\
	&	1300	&		26				&	90	&	16	\\
12.89+0.49	&	20	&	$	30	\pm	3.00	$	&	49	&	17	\\
	&	40	&	$	360	\pm	36.00	$	&	49	&	17	\\
	&	59	&	$	1100	\pm	110.00	$	&	49	&	17	\\
	&	65	&	$	1399.45	\pm	191.26	$	&	600x60	&	4	\\
	&	90	&	$	1306.75	\pm	98.75	$	&	600x90	&	4	\\
	&	101	&	$	2200	\pm	220.00	$	&	49	&	17	\\
	&	135	&	$	2370	\pm	237.00	$	&	49	&	17	\\
	&	140	&	$	1179.40	\pm	321.37	$	&	700x100	&	4	\\
	&	180	&	$	2370	\pm	237.00	$	&	49	&	17	\\
	&	350	&	$	340	\pm	68.00	$	&	120	&	8	\\
	&	400	&	$	210	\pm	21.00	$	&	49	&	17	\\
	&	450	&	$	200	\pm	50.00	$	&	18	&	18	\\
W33A	&	20	&	$	113	\pm	11.00	$	&	49	&	19	\\
	&	25	&	$	268	\pm	21.00	$	&	300x45	&	2	\\
	&	33	&	$	539	\pm	36.00	$	&	6.8	&	20	\\
	&	40	&	$	1000	\pm	100.00	$	&	49	&	8	\\
	&	42	&	$	1300	\pm	130.00	$	&	60	&	21	\\
	&	59	&	$	2350	\pm	235.00	$	&	49	&	8	\\
	&	73	&	$	3400	\pm	340.00	$	&	60	&	21	\\
	&	77	&	$	4100	\pm	410.00	$	&	60	&	21	\\
	&	101	&	$	4050	\pm	405.00	$	&	49	&	8	\\
	&	135	&	$	4000	\pm	400.00	$	&	60	&	2	\\
	&	180	&	$	2750	\pm	275.00	$	&	49	&	8	\\
	&	350	&	$	960	\pm	190.00	$	&	120	&	8	\\
	&	400	&	$	300	\pm	30.00	$	&	49	&	7	\\
	&	450	&	$	240	\pm	60.00	$	&	18	&	22	\\
	&	850	&	$	45	\pm	9.00	$	&	18	&	22	\\
	&	1000	&	$	41	\pm	8.00	$	&	65	&	23	\\
	&	1300	&		11				&	90	&	14	\\
24.49-0.04	&	12	&	$	15.5	\pm	2.20	$	&	300x45	&	2	\\
	&	25	&	$	81	\pm	8.10	$	&	300x45	&	2	\\
	&	60	&	$	1476	\pm	88.00	$	&	90x300	&	2	\\
	&	100	&	$	3514	\pm	14.00	$	&	180x300	&	2	\\
	&	350	&	$	190	\pm	37.00	$	&	120	&	8	\\
W43S	&	12	&	$	218	\pm	20.00	$	&	300x45	&	2	\\
	&	12.5	&		235				&	22	&	24	\\
	&	12.6	&		121				&	2	&	24	\\
	&	19	&		610				&	12	&	24	\\
	&	25	&	$	1697	\pm	136.00	$	&	300x45	&	2	\\
	&	60	&	$	7501	\pm	525.00	$	&	90x300	&	2	\\
	&	65	&	$	3568.90	\pm	1118.72	$	&	600x60	&	4	\\
	&	90	&	$	1721.62	\pm	679.20	$	&	600x90	&	4	\\
	&	100	&	$	11669	\pm	3151.00	$	&	180x300	&	2	\\
	&	140	&	$	3374.24	\pm	335.96	$	&	700x100	&	4	\\
	&	160	&	$	2945.65	\pm	1033.27	$	&	700x150	&	4	\\
	&	350	&	$	440	\pm	88.00	$	&	120	&	8	\\
	&	1300	&		8				&	90	&	14	\\
31.41+0.31	&	25	&	$	52	\pm	5.20	$	&	300x45	&	2	\\
	&	60	&	$	1093	\pm	197.00	$	&	90x300	&	2	\\
	&	65	&	$	1272.43	\pm	176.35	$	&	600x60	&	4	\\
	&	90	&	$	1271.51	\pm	187.27	$	&	600x90	&	4	\\
	&	100	&	$	2815	\pm	394.00	$	&	180x300	&	2	\\
	&	140	&	$	1867.15	\pm	192.69	$	&	700x100	&	4	\\
	&	350	&	$	740	\pm	150.00	$	&	120	&	8	\\
	&	450	&	$	84	\pm	17.00	$	&	9	&	25	\\
	&	850	&	$	27	\pm	1.40	$	&	15	&	25	\\
	&	1350	&	$	4.9	\pm	1.00	$	&	22	&	25	\\
	&	2000	&	$	2.9	\pm	0.60	$	&	34	&	25	\\
W44	&	12	&	$	140.2	\pm	14.02	$	&	300x45	&	2	\\
	&	25	&	$	1106	\pm	110.60	$	&	300x45	&	2	\\
	&	60	&	$	11500	\pm	1150.00	$	&	90x300	&	2	\\
	&	65	&	$	5870.26	\pm	968.91	$	&	600x60	&	4	\\
	&	100	&	$	32460	\pm	3246.00	$	&	180x300	&	2	\\
	&	140	&	$	6928.97	\pm	2480.48	$	&	700x100	&	4	\\
40.50+2.54	&	12	&	$	31.8	\pm	4.50	$	&	300x45	&	2	\\
	&	25	&	$	242	\pm	24.00	$	&	300x45	&	2	\\
	&	60	&	$	2351	\pm	423.00	$	&	90x300	&	2	\\
	&	65	&	$	475.71	\pm	124.26	$	&	600x60	&	4	\\
	&	90	&	$	459.52	\pm	304.48	$	&	600x90	&	4	\\
	&	100	&	$	4218	\pm	840.00	$	&	180x300	&	2	\\
	&	140	&	$	920.73	\pm	84.86	$	&	700x100	&	4	\\
	&	160	&	$	1524.35	\pm	368.28	$	&	700x150	&	4	\\
	&	350	&	$	240	\pm	48.00	$	&	120	&	8	\\
	&	450	&	$	215	\pm	54.00	$	&	18	&	10	\\
	&	850	&	$	33	\pm	7.00	$	&	18	&	10	\\
35.20-0.74	&	12	&	$	4.26	\pm	0.43	$	&	300x45	&	2	\\
	&	60	&	$	1930	\pm	193.00	$	&	90x300	&	2	\\
	&	90	&	$	1929.47	\pm	60.96	$	&	600x90	&	4	\\
	&	100	&	$	1124	\pm	112.40	$	&	180x300	&	2	\\
	&	140	&	$	3046.32	\pm	629.56	$	&	700x100	&	4	\\
	&	160	&	$	4930	\pm	687.84	$	&	700x150	&	4	\\
59.78+0.06	&	12	&	$	14.43	\pm	1.44	$	&	300x45	&	2	\\
	&	25	&	$	108.8	\pm	10.88	$	&	300x45	&	2	\\
	&	60	&	$	982.5	\pm	98.25	$	&	90x300	&	2	\\
	&	100	&	$	1631	\pm	163.10	$	&	180x300	&	2	\\
ON1	&	12	&	$	1.1	\pm	0.10	$	&	300x45	&	2	\\
	&	25	&	$	58.8	\pm	4.70	$	&	300x45	&	2	\\
	&	60	&	$	1431	\pm	115.00	$	&	90x300	&	2	\\
	&	65	&	$	1216.1	\pm	99.83	$	&	600x60	&	4	\\
	&	90	&	$	928.24	\pm	70.58	$	&	600x90	&	4	\\
	&	100	&	$	3119	\pm	312.00	$	&	180x300	&	2	\\
	&	160	&	$	2704.73	\pm	138.89	$	&	700x150	&	4	\\
	&	350	&	$	650	\pm	130.00	$	&	120	&	8	\\
ON2S	&	12	&	$	74	\pm	4.50	$	&	300x45	&	2	\\
	&	25	&	$	481	\pm	29.00	$	&	300x45	&	2	\\
	&	60	&	$	5446	\pm	545.00	$	&	90x300	&	2	\\
	&	100	&	$	<	6985		$	&	180x300	&	2	\\
	&	350	&	$	510	\pm	100.00	$	&	120	&	8	\\
	&	1300	&		9				&	90	&	16	\\
W75N	&	12	&	$	44.91	\pm	4.49	$	&	300x45	&	2	\\
	&	25	&	$	757.8	\pm	75.78	$	&	300x45	&	2	\\
	&	60	&	$	12160	\pm	1216.00	$	&	90x300	&	2	\\
	&	100	&	$	15950	\pm	1595.00	$	&	180x300	&	2	\\
W75OH	&	65	&	$	1753.16	\pm	266.68	$	&	600x60	&	4	\\
	&	90	&	$	1398.82	\pm	346.79	$	&	600x90	&	4	\\
	&	140	&	$	6880.15	\pm	315.18	$	&	700x100	&	4	\\
	&	160	&	$	12253.6	\pm	2280.54	$	&	700x150	&	4	\\
S140	&	12	&	$	332	\pm	40.00	$	&	30	&	26	\\
	&	20	&	$	740	\pm	185.00	$	&	3.5	&	26	\\
	&	25	&	$	1694	\pm	170.00	$	&	30	&	26	\\
	&	35	&	$	5700	\pm	1425.00	$	&	34	&	26	\\
	&	53	&	$	8200	\pm	2050.00	$	&	17	&	26	\\
	&	62	&	$	7600	\pm	130.00	$	&	49	&	26	\\
	&	76	&	$	9200	\pm	150.00	$	&	49	&	26	\\
	&	80	&	$	9900	\pm	2475.00	$	&	37	&	26	\\
	&	101	&	$	7700	\pm	150.00	$	&	49	&	26	\\
	&	111	&	$	7500	\pm	150.00	$	&	49	&	26	\\
	&	162	&	$	4700	\pm	120.00	$	&	49	&	26	\\
	&	175	&	$	54001	\pm	350.00	$	&	45	&	26	\\
	&	350	&	$	1210	\pm	240.00	$	&	120	&	8	\\
	&	400	&	$	3508	\pm	8.00	$	&	49	&	26	\\
	&	1300	&	$	1.4	\pm	0.25	$	&	30	&	22	\\
CepA	&	12	&	$	170	\pm	60.00	$	&	300x45	&	2	\\
	&	25	&	$	860	\pm	215.00	$	&	300x45	&	2	\\
	&	50	&	$	10600	\pm	2650.00	$	&	20	&	27	\\
	&	60	&	$	17000	\pm	3400.00	$	&	90x300	&	2	\\
	&	65	&	$	6375.83	\pm	450.36	$	&	600x60	&	4	\\
	&	100	&	$	230004	\pm	600.00	$	&	180x300	&	2	\\
	&	125	&	$	33100	\pm	9170.00	$	&	50	&	28	\\
	&	140	&	$	7456.13	\pm	401.77	$	&	700x100	&	4	\\
	&	160	&	$	10029.3	\pm	1295.15	$	&	700x150	&	4	\\
	&	350	&	$	1500	\pm	300.00	$	&	120	&	8	\\
	&	400	&	$	2570	\pm	741.00	$	&	50	&	28	\\
	&	450	&	$	737	\pm	140.00	$	&	20	&	29	\\
	&	800	&	$	86	\pm	10.00	$	&	20	&	29	\\
	&	1300	&	$	26	\pm	8.00	$	&	40	&	30	\\
NGC7538-IRS9	&	12.5	&	$	74	\pm	13.00	$	&	9	&	31	\\
	&	20	&	$	124	\pm	30.00	$	&	6	&	31	\\
	&	25	&	$	260	\pm	50.00	$	&	6	&	31	\\
	&	30	&	$	500	\pm	150.00	$	&	40	&	31	\\
	&	50	&	$	1300	\pm	390.00	$	&	40	&	31	\\
	&	100	&	$	2700	\pm	810.00	$	&	55	&	31	\\
	&	350	&	$	330	\pm	66.00	$	&	120	&	8	\\
	&	1000	&	$	51	\pm	5.00	$	&	55	&	31	\\
S157	&	12	&	$	29	\pm	3.00	$	&	300x45	&	2	\\
	&	25	&	$	233	\pm	12.00	$	&	300x45	&	2	\\
	&	60	&	$	1759	\pm	123.00	$	&	90x300	&	2	\\
	&	100	&	$	264	\pm	303.00	$	&	180x300	&	2	\\
	&	350	&	$	280	\pm	56.00	$	&	120	&	8	\\
	&	850	&	$	5.9	\pm	1.20	$	&	18	&	10	
\enddata

{$^1$REFERENCES: (1) MSX Database. (2) IRAS PSC 1988. (3) Mottram et al. 2010. (4) AKARI Database. (5) SCUBA Legacy Catalogs. (6) Aguirre et al. 2011. (7) Mookerjea et al. 1999. (8) Mueller et al. 2002. (9) Dent, Matthews, \& Ward-Thompson 1998. (10) Jenness, Scott, \& Padman 1995. (11) McCutcheon et al. 1995. (12) Chini, Henning, \& Pfau 1991. (13) Tothill 1999. (14) Chini et al. 1986a. (15) Fazio et al. 1978. (16) Chini et al. 1986b. (17) Jaffe et al. 1984. (18) Jenness 1996. (19) Evans et al. 1979. (20) Dyck \& Simon 1977. (21) Stier et al. 1984. (22) Guertler et al. 1991. (23) Cheung et al. 1980. (24) Soifer \& Pipher 1975. (25) Hatchell et al. 2000. (26) Zhou et al. 1994. (27) Ellis et al. 1990. (28) Evans et al. 1981b. (29) Moriarty-Schieven, Snell, \& Hughes 1991. (30) Gordon 1990. (31) Werner et al. 1979.}
\end{deluxetable}

\begin{deluxetable}{ccccccc}
\tablecolumns{7}
\tabletypesize{\footnotesize}
\tablecaption{Bayesian linear correlations \label{tab5}}
\tablewidth{0pt} 
\tablehead{
\colhead{Line} &
\colhead{Set} &
\colhead{Variables} &
\colhead{Cutoff} &
\colhead{b} &
\colhead{a} &
\colhead{$\sigma_{int}$}
}
\startdata
HCO$^+$ (3-2)	&	Galactic	&	L$_{bol}$-L$'$	&	None	&	$	1.10	\pm	0.12	$	&	$	3.26	\pm	0.16	$	&	0.15	\\
HCO$^+$ (3-2)	&	Extragalactic	&	L$_{bol}$-L$'$	&	None	&	$	0.81	\pm	0.21	$	&	$	5.22	\pm	1.68	$	&	0.06	\\
HCO$^+$ (3-2)	&	Combined	&	L$_{bol}$-L$'$	&	None	&	$	1.04	\pm	0.02	$	&	$	3.33	\pm	0.07	$	&	0.12	\\
HCN (1-0)	&	Galactic	&	L$_{bol}$-L$'$	&	None	&	$	1.34	\pm	0.07	$	&	$	2.27	\pm	0.14	$	&	0.17	\\
HCN (1-0)	&	Extragalactic	&	L$_{bol}$-L$'$	&	None	&	$	1.07	\pm	0.06	$	&	$	2.31	\pm	0.50	$	&	0.04	\\
HCN (1-0)	&	Combined	&	L$_{bol}$-L$'$	&	None	&	$	1.02	\pm	0.01	$	&	$	2.76	\pm	0.08	$	&	0.15	\\
HCN (1-0)	&	Galactic	&	L$_{bol}$-L$'$	&	$>$10$^3$ $\lsun$	&	$	1.17	\pm	0.11	$	&	$	2.64	\pm	0.22	$	&	0.17	\\
HCN (1-0)	&	Combined	&	L$_{bol}$-L$'$	&	$>$10$^3$ $\lsun$	&	$	0.99	\pm	0.01	$	&	$	2.99	\pm	0.08	$	&	0.11	\\
HCN (1-0)	&	Galactic	&	L$_{bol}$-L$'$	&	$>$10$^{4.5}$ $\lsun$	&	$	0.88	\pm	0.15	$	&	$	3.40	\pm	0.35	$	&	0.17	\\
HCN (1-0)	&	Combined	&	L$_{bol}$-L$'$	&	$>$10$^{4.5}$ $\lsun$	&	$	0.94	\pm	0.02	$	&	$	3.36	\pm	0.11	$	&	0.17	\\
\tableline																					
HCO$^+$ (3-2)	&	BGPS-MSX	&	L$'$-M$_{iso}$	&	None	&	$	0.84	\pm	0.17	$	&	$	-1.62	\pm	0.56	$	&	0.14	\\
HCO$^+$ (3-2)	&	H$_2$O maser	&	L$'$-M$_{iso}$	&	None	&	$	0.85	\pm	0.18	$	&	$	-1.82	\pm	0.70	$	&	0.09	\\
HCO$^+$ (3-2)	&	Combined	&	L$'$-M$_{iso}$	&	None	&	$	0.78	\pm	0.10	$	&	$	-1.48	\pm	0.35	$	&	0.08	\\
HCO$^+$ (3-2)	&	BGPS-MSX	&	L$'$-M$_{vir}$	&	None	&	$	1.06	\pm	0.43	$	&	$	-2.23	\pm	1.34	$	&	0.32	\\
HCO$^+$ (3-2)	&	H$_2$O maser	&	L$'$-M$_{vir}$	&	None	&	$	1.00	\pm	0.20	$	&	$	-1.86	\pm	0.63	$	&	0.12	\\
HCO$^+$ (3-2)	&	Combined	&	L$'$-M$_{vir}$	&	None	&	$	1.06	\pm	0.18	$	&	$	-2.13	\pm	0.56	$	&	0.15	
\enddata
\end{deluxetable}

\clearpage

\begin{figure}
\centering
\epsscale{0.80}
\plotone{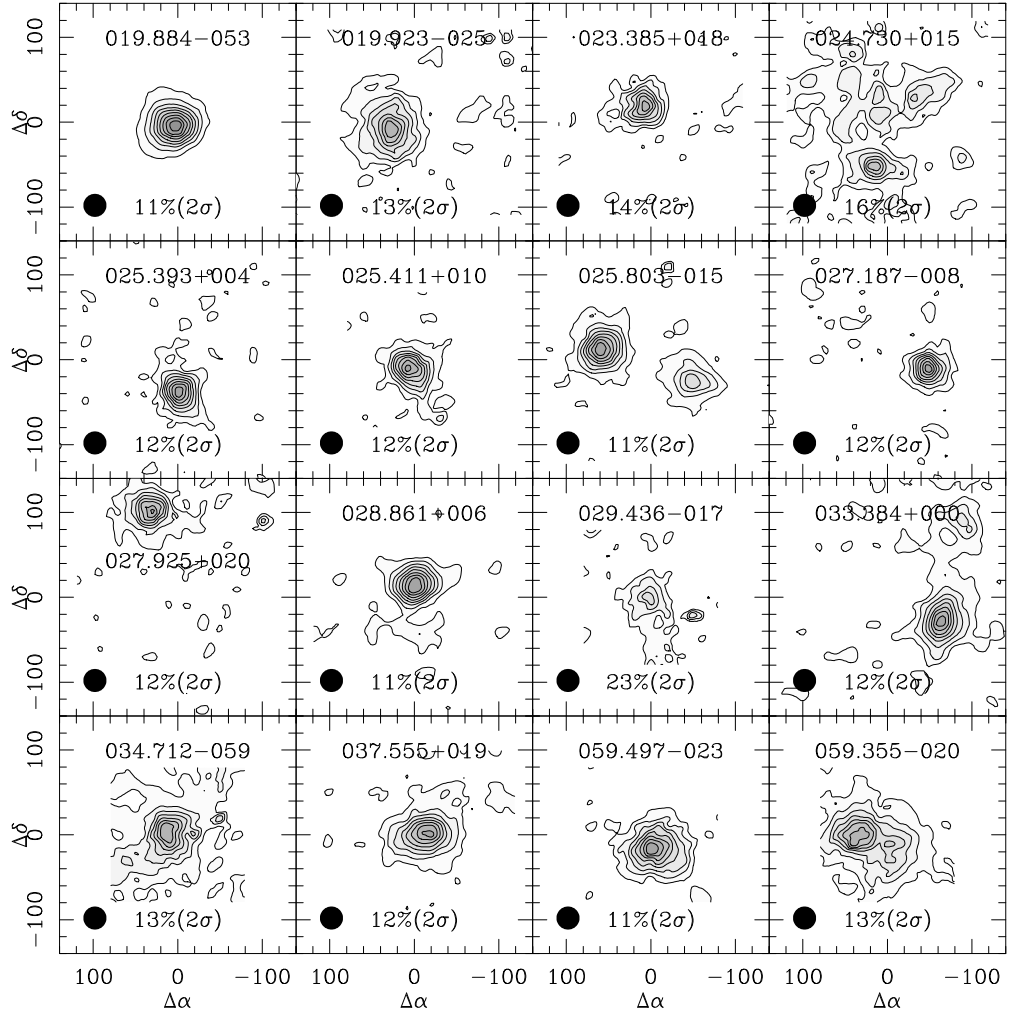}
\caption{HCO$^+$ (3-2) maps for BGPS/RMS sources. The lowest intensity contour and the contour interval are both 2 $\sigma$ in each map. The percentage of the 2 $\sigma$ intensity is also specified.}
\end{figure}

\begin{figure}
\centering
\plotone{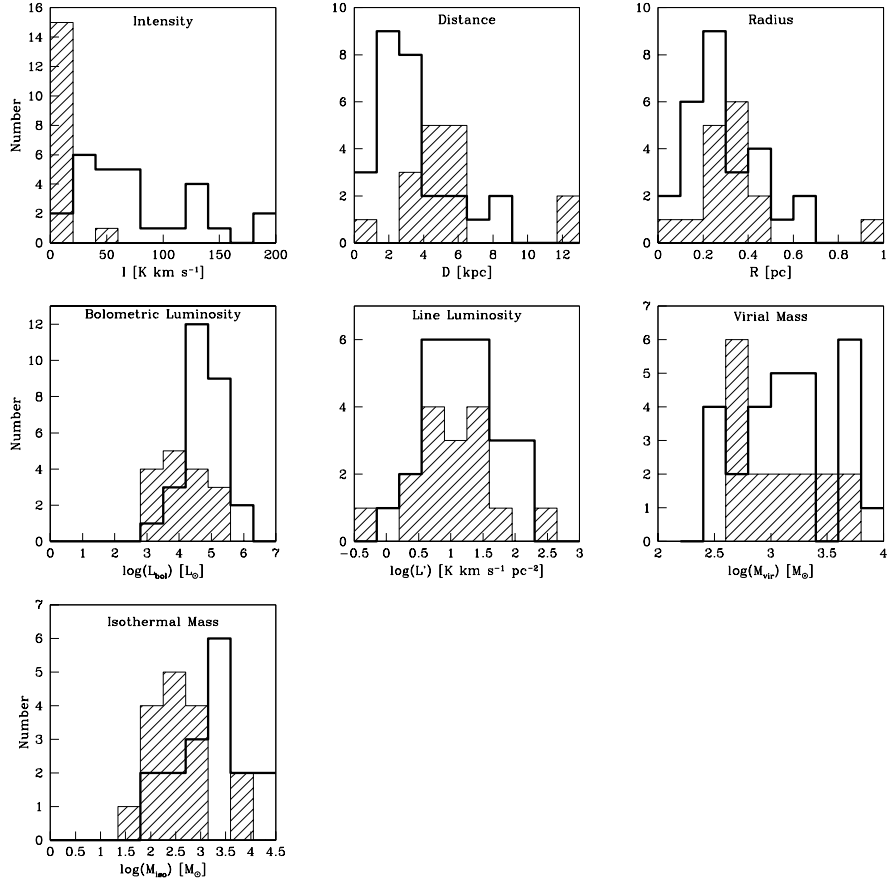}
\caption{Histograms of source properties. The shaded histograms represent the BGPS/RMS sources and the bold line histograms represent the water-maser sources.}
\end{figure}

\begin{figure}
\centering
   \vspace*{5in}
   \leavevmode
   \includegraphics{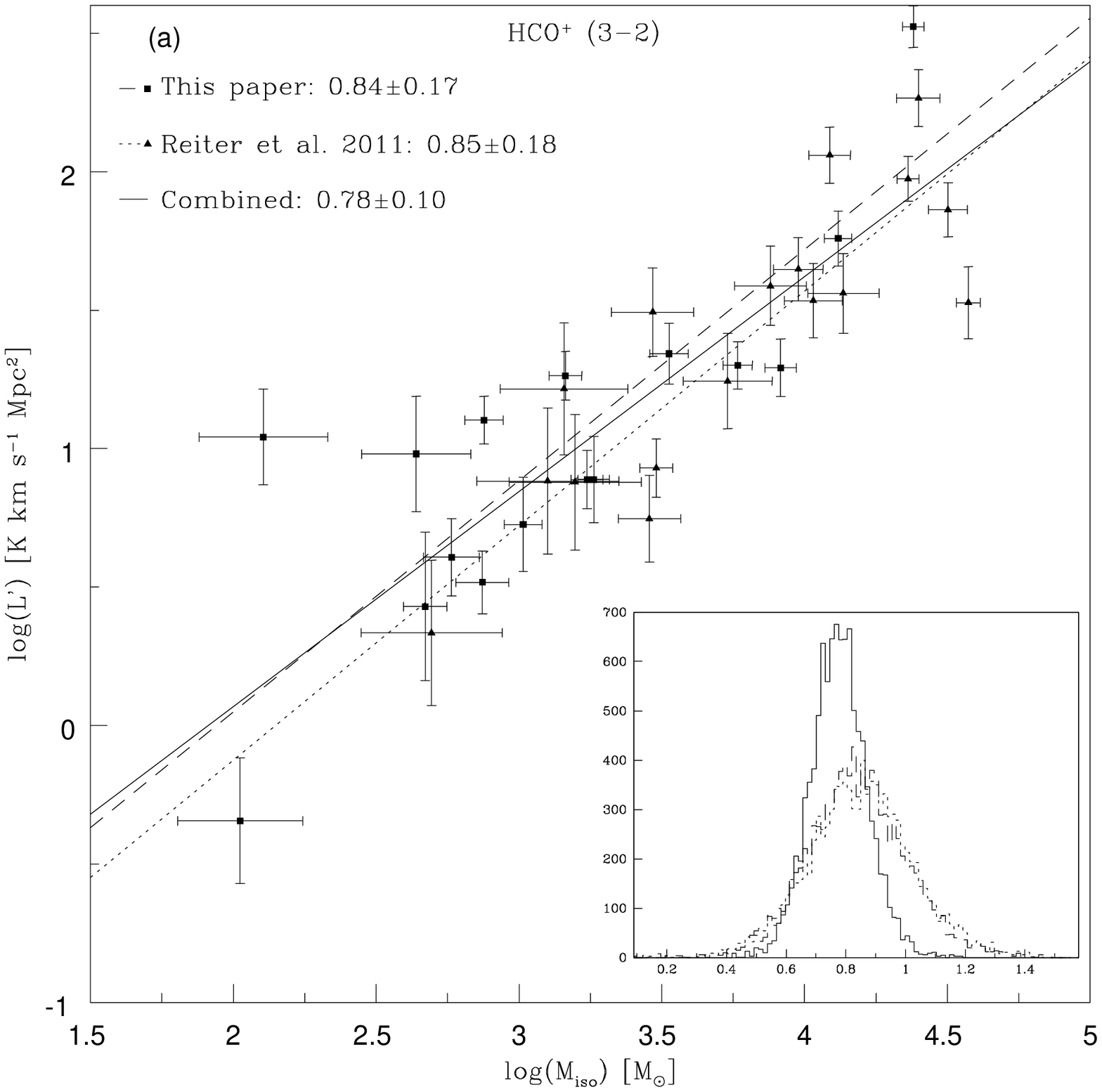}
   \includegraphics{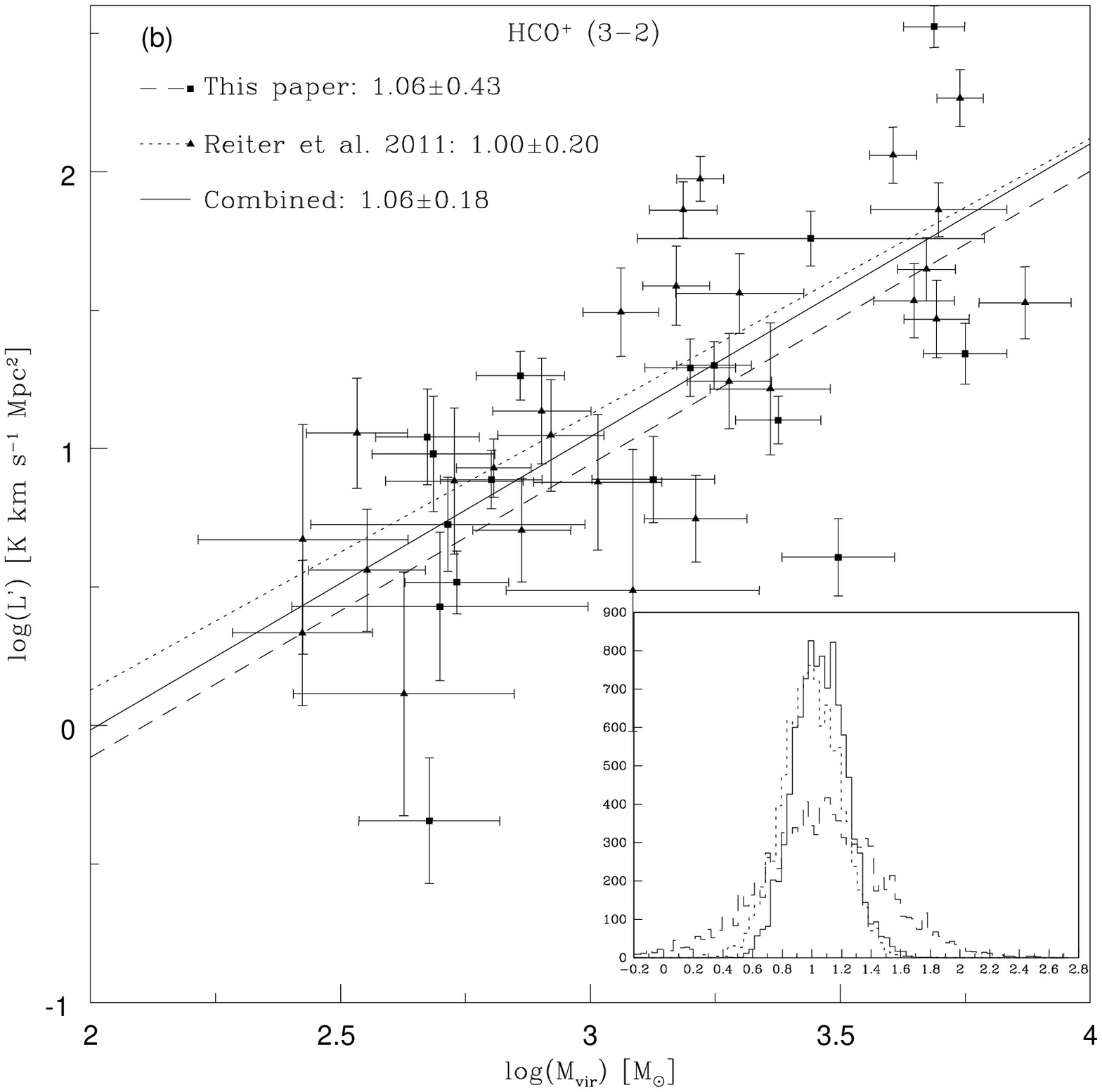}
   \includegraphics{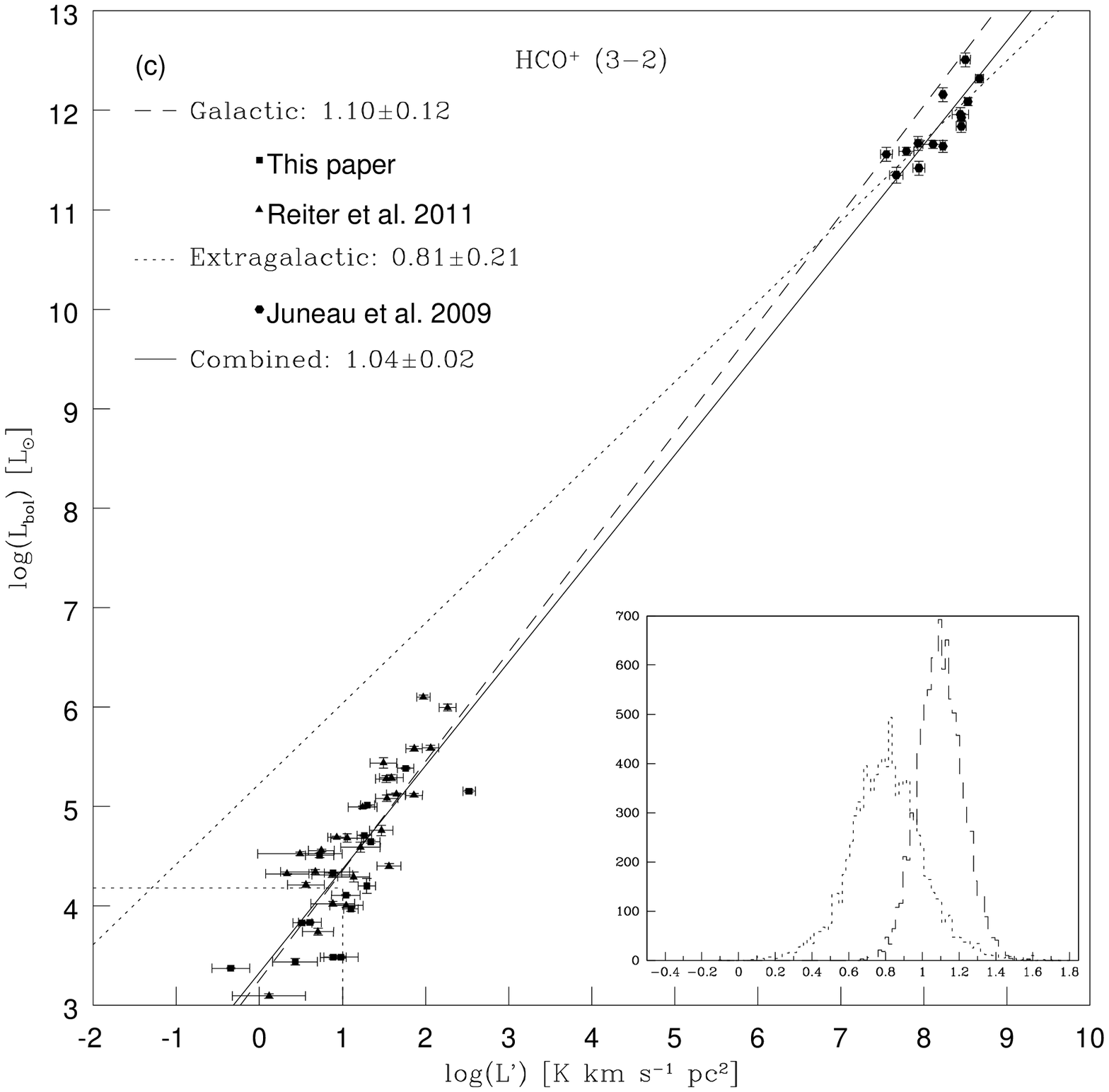}
   \includegraphics{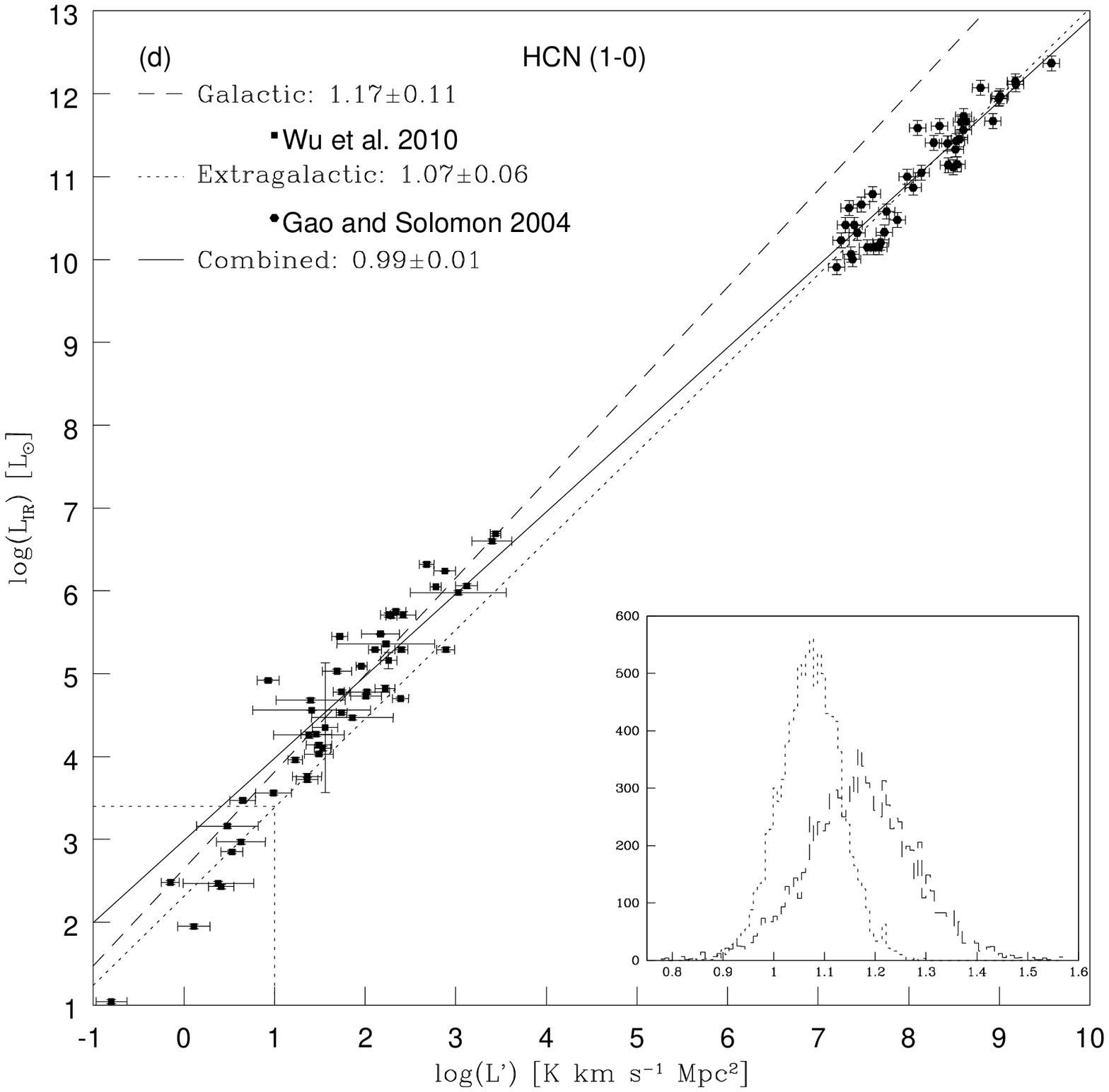}
 \vskip 3.0in
\caption{(a) log(L$'$)-log(M$_{iso}$) for HCO$^+$ (3-2). (b) log(L$'$)-log(M$_{vir}$) for HCO$^+$ (3-2). (c) log(L$_{bol}$)-log(L$'$) for HCO$^+$ (3-2). (d) log(L$_{bol}$)-log(L$'$) for HCN (1-0). The inset histograms display the range of slopes determined from the Bayesian routine.}
\end{figure}

\end{document}